\documentclass[12pt]{JHEP3}
\usepackage[centertags]{amsmath}
\usepackage{amssymb}
\newcommand{\half}{\frac{1}{2}}

 \newcommand{\m}{{\mu}}
 \newcommand{\n}{{\nu}}
 \newcommand{\ep}{{\epsilon}}

 \newcommand{\D}{{\Delta}}

 \newcommand{\ch}{{\chi}}

 \newcommand{\p}{\partial}
 \newcommand{\pslash}{\p \llap{/}}

 \newcommand{\ra}{\rightarrow}

\title{Indices for Superconformal Field Theories in $3,
5$ and $ 6$ Dimensions}

\author{Jyotirmoy Bhattacharya$^{a}$,
Sayantani Bhattacharyya$^{a}$,
Shiraz Minwalla$^{a}$ and 
Suvrat Raju$^{a,b}$ \\
\small{\emph{$^{a}$Department of Theoretical Physics,
                   Tata Institute of Fundamental Research,}}\\
\small{\emph{Homi Bhabha Rd, Mumbai 400005}}  \\
\small{\emph{$^{b}$Department of Physics, Harvard University, Cambridge MA 02138, USA}}  }

\abstract{
We present a trace formula for a Witten type Index
for superconformal field theories in $d=3,5$ and $6$ dimensions,
generalizing a similar recent construction in $d=4$. We perform a
detailed study of the decomposition of long representations into
sums of short representations at the unitarity bound to demonstrate
that our trace formula yields the most general index (i.e. quantity
that is guaranteed to be protected by superconformal symmetry alone)
for the corresponding superalgebras. Using the dual gravitational
description, we compute our index for the theory on the world volume
of $N$ M2 and M5 branes in the large $N$ limit. We also compute our 
index for recently constructed Chern Simons theories in three 
dimensions in the large $N$ limit, and find that, in certain cases, 
this index undergoes a large $N$ phase transition as a function of 
chemical potentials.  
}

\preprint{
  TIFR/TH/08-01 \\
    HUTP-08/A0001 \\
    \texttt{arXiv:0801.1435 [hep-th]}
}

\begin{document}

\maketitle
\section{Introduction}

Supersymmetric fixed points of the renormalization group equations are 
believed to be always either free or superconformally invariant. 
Thus the IR/UV behaviour of any supersymmetric field theory, if 
nontrivial, is governed by a superconformal fixed point. 
Consequently, the study of superconformal dynamics has a special 
place in the study of supersymmetric field theories.  

In radial quantization, the 
 Hilbert space of any unitary superconformal field theory may be 
decomposed into a direct sum over irreducible unitary, lowest energy \
representation of the superconformal algebra. Such 
representations have been classified 
in every dimension (see \cite{Dobrev:1985qv,Dobrev:1985vh,Dobrev:1985qz,Minwalla:1997ka,Dobrev:2002dt} and references therein); the list of these representations turn 
out to include a special set of BPS representations. These  
representations are called `short' because they have fewer states than 
generic representations (we explain this more precisely below);
they also have the property that the energies of all states they host are 
determined by the other conserved charges that label the representation. 

Consider any fixed line of superconformal field theories labeled by some 
continuous `coupling constant' $\lambda$. Suppose that, at any given value of 
$\lambda$, the Hilbert space of the theory possesses some states that 
transform in short representations of the superconformal algebra. Under an 
infinitesimal variation of $\lambda$ the energies of the corresponding 
states can only change if some of these representations jump from being 
short to long. However short representation always contain fewer states 
than long representations with (almost) the same quantum numbers. 
As a consequence, the jump of a single BPS representation from short to long 
is inconsistent with the continuity of the spectrum of the theory
as a function of $\lambda$. Indeed such jumps are consistent with continuity 
only when they occur simultaneously for a group of short representations 
that have the property that their state content is identical to the content of 
a long representation. Such a bunch of BPS representations can continuously 
be transmuted into a long representation, after which the energies of its 
constituent states can be renormalized.

Consequently, a detailed study of 
all possible ways in which short representations can combine up into long 
representations permits the the classification 
of superconformal indices for superconformal field theories.\footnote{
By a superconformal index we mean any function of the spectrum that 
is forced by the superconformal algebra to remain constant under continuous 
variations of the spectrum.} In this paper we perform this study
 for superconformal algebras in $d=3, 5$ and $6$ and use 
our results to provide a complete classification of all 
superconformal indices in these dimensions. In each of these
cases, we also provide a trace formula that, when evaluated in a 
superconformal field theory, may be used to extract all these superconformal
indices. This is the analogue of the trace formula described in \cite{Witten:1982df} for the Witten index.
Thus the Witten index we define in this paper constitutes the most 
general superconformal Index in $d=3, 5, 6$.\footnote{The corresponding 
results are already known in $d=4$ \cite{Kinney:2005ej}. In 2 dimensions
the analogue of the indices we will study here is the famous `elliptic genus' \cite{Lerche:1987qk,Pilch:1986en} while superconformal algebras do not exist in $d>6$. }

We then proceed to compute our superconformal Witten Index for specific 
superconformal field theories. We first perform this computation for the 
superconformal field theories on the world volume of $N$ M2
and $N$ M5 branes, at $N=1$ (using field theory) and at $N=\infty$ (using 
the dual supergravity description). We find that our index has significant 
cancellations compared to the simple partition function over supersymmetric 
states.  In each case, the density of states in the Index grows slower in 
comparison to the supersymmetric entropy. 
We also compute our index
for some of the Chern Simons superconformal field theories 
recently analyzed by Gaiotto and Yin \cite{Gaiotto:2007qi}; and find that, in some cases, 
this index undergoes a large $N$ phase transition as a function of 
chemical potentials. 

This paper is divided into 3 self-contained parts. Superconformal algebras in $d=3$ are analyzed in Section \ref{d3section}, in $d=6$ are discussed in Section \ref{dsamansix} and in $d=5$ are discussed in \ref{d5section}. In each section, we describe the relevant algebra and its unitary representations. We then discuss short representations and enumerate all possible ways in which short representations can pair up into long representations. We use this enumeration to provide, in each dimension, an exhaustive list of all indices that are protected by group theory alone. We also provide a trace formula for a Witten type index that may be evaluated via a path integral. These indices count states that are annihilated by a particular supercharge. We discuss how the Witten Index may be expanded out in characters of the subalgebra of the superconformal algebra that commutes with this supercharge. The coefficients of these characters in the Witten Index are nothing but the indices mentioned above. 

In $d=3$, we evaluate our index in three different theories: (a) Supergravity on $AdS_4 \times S^7$ (b) the worldvolume theory of a single $M2$ brane and (c) the Chern-Simons matter theories recently discussed in \cite{Gaiotto:2007qi}. In $d=6$, we evaluate our index in supergravity on $AdS_7 \times S^4$ and in the worldvolume theory of a single $M5$ brane. 

Finally, we wish to mention a subtlety that we have avoided in our discussion above. Indices may fail to be protected if the spectrum of the theory contains a continuum \cite{Witten:1982df, Cecotti:1992qh} or is singular for some parameters. Lately, this issue has attracted interest in the context of 2 dimensional conformal field theories and we direct the reader to \cite{Denef:2007vg,Sen:2007qy,Raju:2007uj} for some recent discussions. 

\section{d=3}
\label{d3section}
\subsection{The Superconformal Algebra
 and its Unitary Representations}
The bosonic subalgebra of the $d=3$ superconformal algebra is  
$SO(3,2) \times SO(n)$ (the conformal algebra times the R symmetry algebra). 
The anticommuting generators in this algebra may be divided into 
the generators of supersymmetry ($Q)$ and the generators of 
superconformal symmetries ($S)$. Supersymmetry generators transform in the 
vector representation of the R-symmetry group $SO(n)$,\footnote{In the literature on the worldvolume theory of the $M2$ brane, the supercharges are taken to transform in a spinor of $SO(8)$. This is consistent with the statement above, because for $n=8$, the vector and spinor representations are related by a triality flip and a change of basis takes one to the other.} have charge half 
under dilatations (the $SO(2)$ factor of the compact $SO(3)\times SO(2)
\in SO(3,2)$) and are spinors under the $SO(3)$ factor of the same 
decomposition. Superconformal generators $S_i^{\mu} = (Q^i_{\mu})^{\dagger}$ 
transform in the spinor representation of $SO(3)$, have scaling 
dimension (dilatation charge) $(-\half)$, and also transform in the vector 
representation of the R-symmetry group. In our 
notation for supersymmetry generators $i$ is an $SO(3)$ spinor index 
while $\mu$ is an $R$ symmetry vector index.

We pause to remind the reader of the structure of the commutation relations 
and irreducible unitary representations of the $d=3$ superconformal algebra
(see \cite{Minwalla:1997ka} and references therein ). 
As usual, the anticommutator between two supersymmetries 
is proportional to momentum times an $R$ symmetry delta function, and the 
anticommutator between two superconformal generators is obtained by 
taking the Hermitian conjugate of these relations. The most interesting 
relationship in the algebra is the anticommutator between $Q$ and $S$. 
Schematically
\begin{equation*}\{S_i^{\mu},Q^j_{\nu}\} \sim \delta^{\mu}_{\nu}T^{j}_{i} -
\delta^{j}_{i}M^{\mu}_{\nu}\end{equation*} Here $T^{j}_{i}$ are the $U(2) \sim
SO(3)\times SO(2)$ generators and $M^{\mu}_{\nu}$ are the $SO(n)$
generators. 

Irreducible unitary lowest energy representations of this algebra 
possess a distinguished set of lowest energy states called primary states. 
Primary states have the lowest value of $\epsilon_0$ -- the eigenvalue of the 
dilatation (or energy) operator -- of all states in their representation. 
They transform in irreducible representation of $SO(3) \times SO(n)$, 
and are annihilated by all superconformal generators and special conformal 
generators.\footnote{i.e. all generators of negative scaling 
dimension.}

Primary states are special because all other states in the unitary (always 
infinite dimensional) representation may be obtained by acting on the 
primary with the generators of supersymmetry and momentum. For a primary with energy $\epsilon_0$, a state obtained 
by the action of $k$ different $Q$ s on the primary has energy 
$\epsilon_0+{k \over 2}$, and is said to be a state at the $k^{\rm th}$ level in the 
representation. The energy, $SO(3)$ highest weight 
(denoted by $j=0, \half, 1 \ldots$)  and the R-symmetry highest weights 
($h_1, h_2 \ldots h_{[n/2]})$ \footnote{$h_i$ are eigenvalues under rotations
in orthogonal 2 planes in $R^n$. Thus, for instance, $\{h_i \}=(1,0,0, ..0)$ 
in the vector representation} of primary states form a complete set 
of labels for the entire representation in question.

Any irreducible representation of the superconformal algebra 
may be decomposed into a finite number of distinct irreducible 
representations  of the conformal algebra. The latter are labeled 
by their own primary states, which have a definite lowest energy 
and transform in a given irreducible representation of $SO(3)$. 
The state content of an irreducible representation of the superconformal 
algebra is completely specified by the quantum numbers of its 
constituent conformal primaries.

As we have mentioned in the introduction, the superconformal algebra possesses
special short or BPS representations which we will now explore in more 
detail. Consider a representation of the algebra, whose primary transforms 
in the spin $j$ representation of $SO(3)$ and in the $SO(n)$ 
representation labeled by highest weights 
$\{h_i\}\ i = 1,\cdots,\left[\frac{n}{2}\right]$. We normalize primary states to have 
unit norm. The superconformal algebra -- plus the Hermiticity relation 
$(Q^i_\mu)^\dagger= S_{i}^{\mu}$ -- completely determines the inner 
products between any two states in the representation. All states in 
an unitary representation must have positive norm: however this requirement 
is not algebraically automatic, and, in fact imposes a nontrivial restriction 
on the quantum numbers of primary states. This restriction takes the form 
$\epsilon_0 \geq f(j, h_i)$ as we will now explain.\footnote{These techniques have been used in the investigation of unitarity bounds for conformal and superconformal algebras in \cite{Mack:1975je,Dobrev:1985qv,Dobrev:1985vh,Dobrev:1985qz,Minwalla:1997ka,Dolan:2002zh}.}

Let us first consider descendant states,  at level one,  of a representation 
 whose primary has $SO(3)$ and $SO(n)$ 
quantum numbers $j, (h_1 \ldots h_{[n/2]} )$.  It is easy to compute the norm 
of all such states by using the commutation relations of the algebra. 
As long as $j \neq 0$ it turns out that the level one states with lowest norm transform in  in the spin $j-\half$
representation of the conformal group and in the  
$( h_1 +1,\{h_i\}) \ i =
2,\cdots,\left[\frac{n}{2}\right] $ representation of $SO(n)$ \cite{Minwalla:1997ka}. The highest 
weight state in this representation may be written explicitly as (see \cite{Dolan:2002zh})
\begin{equation} \label{ps} |Zn_1\rangle =
A^-_1|h.w\rangle \equiv \left(Q^{-\half}_{1} -
Q^{\half}_1J_-\left(\frac{1}{2J_z}\right)\right)|h.w\rangle
\end{equation}
where $J_-$ denotes the spin lowering operator of $SO(3)$ and 
$Q^{\pm \half}_{1}$ are supersymmetry operators with $j=\pm \half$ and 
$(h_1, h_2, \ldots h_{[n/2]})=(1, 0, \ldots , 0)$. Here $ |h.w\rangle$ is 
a highest weight state with energy $\epsilon_0$, $SU(2)$ charge $j$ and $SO(n)$
charge $(h_1,h_2,\dots,h_{[n/2]})$.
The norm of this state is easily computed and is given by,
\begin{equation}
\langle Zn_1 | Zn_1 \rangle = \left( 1 + \frac{1}{2j} \right)(\epsilon_0-j-h_1-1)
\end{equation}
It follows that the non negativity of norms of states at level one (and so the unitarity of the representation)
requires that the charges of the primary should satisfy
 \begin{equation} \label{lob} \epsilon_0 \geq j+h_1+1 \end{equation}
For $j\neq 0$ this inequality turns out to be the necessary and 
sufficient condition for a representation to be unitary. 

When the primary saturates the bound \eqref{lob} the representation 
possess zero norm states: however it turns out to be consistent to define a 
truncated representation by simply deleting all zero norm states. This 
procedure yields a physically acceptable representation whose quantum numbers 
saturate \eqref{lob}. This truncated representation is 
unitary (has only positive norms) but has fewer states than the generic 
representation, and so is said to be `short' or BPS. 

The set of zero norm states we had to delete, in order to obtain the BPS
representation described above, themselves transform in a 
representation of the superconformal algebra. This representation is 
 labeled by the primary state 
$|Zn_1\rangle$  (see \eqref{ps}).

Let us now turn to the special case $j=0$. In this case $|Zn_1\rangle$ is ill
defined and does not exist; no states with its quantum numbers occur at level 
one. The states of lowest norm at level one transform in the 
spin half $SO(3)$ representation, and have $SO(n)$ highest weights 
$h_1' = h_1 +1,\{h_i\}\ i = 2,\cdots,\frac{n}{ 2}$. The highest weight state
in this representation is  
$|Zn_2\rangle = A^+_1|h.w.\rangle \equiv Q^{\half}_1|h.w\rangle$.
The norm of this state is $(\epsilon_0-h_1)$. Unitarity thus imposes the 
constraint $\epsilon_0 \geq h_1$. However, in this case, this condition is 
necessary but not sufficient to ensure unitarity, as we now explain. 

As we have remarked above, the state $|Zn_1\rangle = A^-_1|h.w\rangle$ is 
ill defined when $j=0$. However $|s_2\rangle = \left( A^+_1A^-_1 \right)|h.w\rangle 
= Q^{\half}_1Q^{-\half}_1|h.w\rangle$ is well defined even in this situation 
(when $j=0$). The norm of this 
state is easily computed and is given by,\footnote{When $j\neq 0$, the norm of 
$|s_2 \rangle$ had to be proportional to $(\epsilon_0-j-h_1-1)$ simply 
because the norm of $|s_2\rangle$ must vanish whenever 
$|Zn_1\rangle$ is of zero norm.
 The algebra that leads to this 
result is correct even at $j=0$ (i.e. when $|Zn_1\rangle$ is ill defined).}
\begin{equation}
\langle s_2 | s_2 \rangle =  (\epsilon_0+j-h_1)(\epsilon_0-j-h_1-1).
\end{equation}

It follows that, at $j=0$, the positivity of norm of all states requires 
either that $\epsilon_0 \geq h_1+1$ or that $\epsilon_0 = h_1$. 
This turns out to be the complete 
set of necessary and sufficient conditions for the existence of unitary 
representations. Representations with $j=0$ and $\epsilon_0 = h_1+1$ or 
$\epsilon_0=h_1$ are 
both short.  The representation at $\epsilon_0 = h_1$ is an isolated  short 
representation since there is no representation in the energy gap 
$h_1\leq \epsilon_0\leq (h_1 +1)$; its first zero norm 
state occurs at level one. 
The first zero norm state in the $j=0$ representation at $\epsilon_0=h_1+1$ 
occurs at level 2 and is given by $| s_2 \rangle$.

In summary, short representations occur when the highest weights of the primary state satisfy one of the following conditions \cite{Minwalla:1997ka}.
\begin{equation}\label{ssh} 
\begin{split}
\epsilon_0 =& j + h_1 + 1\ \ \rm{when}\ j \geq 0,\\
\epsilon_0 =& h_1 \ \ \rm{when}\ j = 0.
\end{split}
\end{equation}
The last condition gives an isolated short representation.

\subsection{Null Vectors and Character Decomposition of a Long Representation at the Unitarity Threshold}

As we have explained in the previous subsection, short representations of the 
$d=3$ superconformal algebra are of two sorts. The energy of a `regular' short 
representation is given by $\epsilon_0=j+h_1+1$. The null states of this 
representation transform in an irreducible representation of the algebra. When $j \neq 0$ the 
highest weights of the primary at the head of this null 
irreducible representation is  given in terms 
of the highest weights  of the representation itself 
by $\epsilon_0'=\epsilon_0+\half,
~~~ j'=j-\half,~~~
h_1'=h_1+1,~~~h_i'=h_i$. Note that 
$\epsilon_0'-j'-h_1'-1= \epsilon_0-j-h_1-1=0$, so 
that the null states also transform in a regular short representation. As 
 the union of the ordinary and null states of such a 
short representation is identical 
to the state content of a long representation at the 
edge of the unitarity bound, we conclude that 
\begin{equation}\label{enf} \lim_{\delta \ra 0}\ch[j+h_1+1 +\delta,j,h_1,h_j] =
\ch[j+h_1+1, j,h_1,h_j] + \ch[j+h_1+3/2, j - \half, h_1 + 1,h_j]
\end{equation}
where $\chi[ \epsilon_0, j, h_i]$ denotes the supercharacter of the irreducible representation with energy 
$\epsilon_0$, $SO(3)$ highest weight $j$ and $SO(n)$ highest weights $\{h_i\}$. Note that the $\chi$ s appearing 
on the RHS of \eqref{enf} are the supercharacters corresponding to short representations.

On the other hand when $j=0$ the null states of the regular short 
representation occur at level 2 and are labelled by a primary with 
highest weights $\epsilon_0'=\epsilon_0+1, ~~~j'=0, h_1'=h_1+2, ~~~
h_i'=h_i$. Note in particular that $j'=0$ and $\epsilon_0'-h_1' = 
\epsilon_0-h_1-1=0$. It follows that the null states of this representation 
transform in an isolated short representation, and we conclude 
\begin{equation}\label{eng} 
\lim_{\delta \ra 0}\ch[h_1+1 + \delta,j=0,h_1,h_j] =
\ch[h_1+1,j=0,h_1,h_j] +
\ch[h_1+2,j=0,h_1+2,h_j]
\end{equation}

Recall that isolated short representations are separated from all other 
representations with the same $SO(3)$ and $SO(n)$ quantum numbers by a gap in 
energy. As a consequence it is not possible to `approach' such representations 
with long representations; consequently we have no equivalent of \eqref{eng} 
at energies equal to $h_1+\delta$. 

For use below we define some notation. We will use $c(j, h_i)$ $({\rm with~~} i=1,2,\dots,[\frac{n}{2}])$ to denote a 
regular short representation with $SO(3)$ and $SO(n)$ highest weights $j$,
${h_i}$ respectively and  $\epsilon_0=j+h_1+1$ (when $j \geq 0$). We will also use the symbol 
$c(-\half, h_1,h_j)$ $({\rm with}~h_1 \geq h_2-1)$ to denote the isolated short representation 
with $SO(3)$ quantum number $0$, $SO(n)$ quantum numbers $h_1+1, {h_j}$ (here $j=2,3,\dots,[\frac{n}{2}]$)
respectively and $\epsilon_0=h_1+1$. The utility of this notation will become apparent below.

\subsection{Indices}

The state content of any unitary superconformal quantum field theory may be 
decomposed into a sum of an (in general infinite number of) irreducible 
representations of the superconformal algebra. This state content 
is completely determined by specifying the number of times any given 
representation occurs in this decomposition. Consider any linear combination 
of the multiplicities of short representations. If this linear 
combination evaluates to zero on every collection of representations 
that appears on the RHS of each of \eqref{enf} and \eqref{eng} (for all values 
of parameters), it is said to be an index. It follows from 
this definition that indices are unaffected by all possible pairing up of 
short representations into long representations, and so are invariant under any
deformation of superconformal Hilbert space under which the spectrum evolves 
continuously. We now proceed to list these indices.

\begin{enumerate}
\item
The simplest indices are simply given by the multiplicities 
of representations in the 
spectrum that never appear on the RHS of \eqref{eng} and \eqref{enf} (for any 
values of the quantum numbers of the long representations on the LHS of those 
equations). All such representations are easy to list; they are $SO(3)$ 
singlet isolated representations whose $SO(n)$ quantum number $h_1-|h_2| \leq 1$ where $h_1$ and $h_2$ are both either integers or half integers, and $h_1 \geq |h_2| \geq 0$.

\item
We can also construct indices from linear combinations of the multiplicities 
of 
representations that do appear on the RHS of \eqref{eng} and \eqref{enf}.
The complete list of such linear combinations is given by
\begin{equation}\label{veca} I_{M,\{h_j\}} = \sum_{p=-1}^{M-|h_2|}(-1)^{p+1}n\{c({p 
\over 2},M-p,h_j\},
\end{equation}
 where $n[R]$ denotes the multiplicities of representations of type $R$ 
and the Index label  $M$ is the value of $h_1+2j$ for every regular 
short representation that appears in  the sum above.
Thus $M \geq |h_2|$ and both $M$ and $h_2$ are either integers or half-integers.Also the set $\{h_j\}$ must satisfy 
the condition $h_2 \geq h_3 .....\geq |h_{[\frac{n}{2}]}|$ where all the $h_i$ are either integers or all are half-integers. 
\end{enumerate}

\subsection{Minimally BPS states: distinguished supercharge and commuting superalgebra}

We will now describe states that are annihilated by at least one supercharge and its conjugate. 
Consider the special supercharge $Q$ with charges $(j=-\half,h_1=1,h_i=0,\ep_0=\half)$. 
Let $S=Q^{\dagger}$; it is easily verified that 
\begin{equation}\label{dela}  \{S, Q \}= \Delta= \ep_0 -(h_1+j)
\end{equation}
Below we will be interested in a  
partition function over states annihilated by $Q$. 
Clearly all such states transform in irreducible representations of that 
subalgebra of the superconformal algebra that commutes with $Q,S$ and 
hence $\Delta$. This subalgebra is easily determined to be a real form of the 
supergroup $D(\frac{n-2}{2},1)$ or $B(\frac{n-3}{2}, 1)$, depending on whether 
$n$ is even or odd. We follow the same notation as \cite{Minwalla:1997ka}.

The bosonic subgroup of this commuting superalgebra is 
$SO(2,1)\times SO(n-2)$. The usual Cartan charge of $SO(2,1)$ (the $SO(2)$ 
rotation) and the Cartan charges of $SO(n-2)$ are  given in terms of the 
Cartan elements of the parent superconformal algebra by 
\begin{equation}\label{repa} E = \ep_0 + j,~~~
H_i=h_{i+1} \ \ \left(\rm{with}\ i=1,2,\dots,[\frac{n-2}{2}]\right).
\end{equation}

\subsection{A Trace formula for the general Index and its Character
Decomposition}

Let us define the Witten index 
\begin{equation}\label{wia} I^{W} = {\rm Tr}_R[(-1)^F \exp(-\beta\Delta + G )],
\end{equation} 
where the trace is evaluated over any Hilbert space $R$  that 
hosts a representation (not necessarily irreducible) of the 
superconformal algebra. Here $F$ is the Fermion number operator; by the 
spin statistics theorem $F=2 j$ in any quantum field theory. $G$ is any element of the subalgebra that commutes with $\{S,Q,\D\}$; by a similarity
transformation, $G$ may be rotated into a linear combination of the
Cartan generators of the subalgebra.

The Witten Index \eqref{wia} receives contributions 
only from states that are annihilated by both $Q$ and $S$ (all other states 
yield contributions that cancel in pairs) and have $\Delta=0$. So, it is independent of $\zeta$. The usual arguments\cite{Witten:1982df} also ensure 
that $I^{W}$ is an index; consequently it must be possible to expand $I^W$
as a linear sum over the indices defined in the previous section. 
In fact it is easy to check 
that for any representation A(reducible or irreducible), 
\begin{equation}\label{wid} I^{W}(A)= \sum_{M,\{h_i\}} I_{M,\{h_i\}}\ch_{sub} (M+2,h_i) +
\sum_{\{h_j\},h_1-|h_2|=0,1}
n\{c(-\half,h_1-1,h_i)\}\ch_{sub}(h_1,h_i).
\end{equation}

where $\ch_{sub}(E,H_i)$ (with $i=1,2,\dots,[\frac{n-2}{2}]$) is the supercharacter of the
subalgebra\begin{footnote}{The supercharacter of a representation $R$ is defined as 
$\chi_{sub}(R) = {\rm tr}_{R} (-1)^{F} {\rm tr}\  e^{{\bf \mu \cdot H}}$, where 
${\bf \mu \cdot H}$ is some linear combination of the Cartan generators specified by a chemical 
potential vector ${\bf \mu}$. $F$ is defined to anticommute with $Q$ and commute with the bosonic 
part of the algebra. The highest weight state is {\em always taken} to have $F=0$.}\end{footnote} 
with $E$ and $H_i$ being the highest weights of a representation of the subalgebra in the convention 
defined by \eqref{repa}. In the first term on the RHS of \eqref{wid} the sum runs over all the 
values of $M,\{h_j\}$ for which $I_{M,\{h_j\}}$ is defined (see below \eqref{veca}). In the second term the sum runs over all
the values of the set $\{h_j\}$ such that $h_2 \geq h_3 .....\geq |h_{[\frac{n}{2}]}|$.
In order to obtain \eqref{wib} we have used 
\begin{equation}\label{wib} I^{W}(c(j,h_1,h_j)) = (-1)^{2j+1}\ch_{sub}(2j+h_1+2,h_i)
\end{equation}
\begin{equation}\label{wic} I^{W}(c(j=-\half,h_1,h_j)) = \ch_{sub}(h_1+1,h_j)
\end{equation}
Equation \eqref{wib} asserts that the set of $\Delta=0$ states 
(the only states that contribute to the Witten Index) in any 
short irreducible representation 
of the superconformal algebra transform in a single irreducible 
representation of the commuting subalgebra. In the case of regular 
short representations, the primary 
of the full representation has $\Delta=1$. The primary of the subalgebra   
is obtained by acting on the primary of the full representation 
with a supercharge with quantum numbers $(j=\half,h_1=1,h_i=0,\ep_0=\half,
 \Delta=-1)$. On the other hand the highest weight of an isolated  
superconformal short primary itself has $\Delta=0$, and so is also a primary  
of the commuting sub super algebra. Equation \eqref{wid} follows immediately from 
these facts.

Note that every index that appears in the list of subsection 2.3 
appears as the coefficient of a distinct subalgebra supercharacter 
in  \eqref{wid}. As supercharacters of distinct irreducible representations 
are linearly independent,  it follows that knowledge of $I^{W}$ is sufficient 
to reconstruct all superconformal indices of the algebra. In this sense  
\eqref{wid} is the most general index that is possible to construct from 
the superconformal algebra alone.

\subsection{The Index over M theory multi gravitons in $AdS_4 \times S^7$}

We will now compute the Witten Index defined above 
in specific examples of three dimensional superconformal field theories. In 
this subsection we focus on the world volume theory of the M2 brane in 
the large $N$ limit. The corresponding theory has  supersymmetries and 16 
superconformal symmetries. The bosonic subgroup of the relevant superconformal 
algebra is $SO(3,2) \times SO(8)$. We take the 
supercharges to transform in the vector representation of $SO(8)$; 
this convention is related to the one used in much of literature on this 
theory by a triality flip.

In the strict large $N$ limit, the index over the M2 brane conformal field 
theory is simply the index over the Fock space of supergravitons for M theory 
on $AdS_4 \times S^7$ \cite{Maldacena:1997re,Aharony:1999ti}. In order to compute this quantity we first compute 
the index over single graviton states; the index over multi gravitons is 
given by the appropriate Bose- Fermi exponentiation (sometimes called the 
plethystic exponential).\footnote{The index we will calculate is sensitive to ${1 \over 16}$ BPS states. However, the ${1 \over 8}$ BPS partition function has been calculated, even at finite $N$, in \cite{Bhattacharyya:2007sa}}

Single particle supergravitons in $AdS_4 \times S^7$ transform in 
an infinite class of representations of the superconformal algebra. 
The primaries for this spectrum have charges (see \cite{Gunaydin:1985tc, biran1984fss})
($\ep_0={n \over 2}$,$j=0$,$h_1={n \over 2}$,$h_2={n \over 2}$,
$h_3={n \over 2}$,$h_4=-{n \over 2}$) ($h_1,h_2,h_3$ and $h_4$ denote 
$SO(8)$ highest weights in the orthogonal basis; recall $Q$s here are taken to transform  in the vector 
rather than the spinor of $SO(8)$) where $n$ runs from $1$ to $\infty$   
(we  are working with the `$U(N)$ theory;  $n=1$ would be omitted 
for the $SU(N)$ theory).  

It is not difficult to decompose each of these irreducible representations of the superconformal 
algebra into representations of the conformal algebra, and thereby compute 
the partition function and the Index over each of these representations. 
The necessary decompositions  were performed in \cite{Gunaydin:1985tc}, 
and we have verified their results independently by means a procedure 
described in in Appendix \ref{racahspeiser}.
The results are listed in 
Table~\ref{tb:d3gravspec} below.\footnote{Some of the conformal representations obtained in 
this decomposition are short (as conformal representations) when $n$ is either 1 or 2; the negative 
contributions in table 1 represent the charges of the null states, which 
physically are operators set to zero by the equations of motion. See \cite{Barabanschikov:2005ri} }

\TABLE{
\caption{d=3 graviton spectrum}
\label{tb:d3gravspec}
\begin{tabular}{|c|c|c|c|c|c|}
	\hline         
range of $n$ & $\epsilon_0[SO(2)]$ & $SO(3)$ & $SO(8)$[orth.(Qs in vector)] & $\Delta$ & contribution  \\
	\hline
        
$n\geq1$   &$\frac{n}{2}$  &$0$          &($\frac{n}{2},\frac{n}{2},\frac{n}{2},\frac{-n}{2}$)                & $0$ & $+$ \\
$n\geq1$   &$\frac{n+1}{2}$&$\half$      &($\frac{n}{2},\frac{n}{2},\frac{n}{2},\frac{-(n-2)}{2}$)            & $0$ & $+$ \\
$n\geq2$   &$\frac{n+2}{2}$&$1$          &($\frac{n}{2},\frac{n}{2},\frac{(n-2)}{2},\frac{-(n-2)}{2}$)        & $0$ & $+$ \\
$n\geq2$   &$\frac{n+3}{2}$&$\frac{3}{2}$&($\frac{n}{2},\frac{(n-2)}{2},\frac{(n-2)}{2},\frac{-(n-2)}{2}$)    & $0$ & $+$ \\
$n\geq2$   &$\frac{n+4}{2}$&   $2$  &($\frac{(n-2)}{2},\frac{(n-2)}{2},\frac{(n-2)}{2},\frac{-(n-2)}{2}$)     &$1$ & $+$ \\
$n\geq2$   &$\frac{n+2}{2}$&$0$         &($\frac{n}{2},\frac{n}{2},\frac{n}{2},\frac{-(n-4)}{2}$)            & $1$ & $+$ \\
$n\geq3$   &$\frac{n+3}{2}$&$\half$      &($\frac{n}{2},\frac{n}{2},\frac{(n-2)}{2},\frac{-(n-4)}{2}$)        & $1$ & $+$ \\
$n\geq3$   &$\frac{n+4}{2}$&$1$          &($\frac{n}{2},\frac{(n-2)}{2},\frac{(n-2)}{2},\frac{-(n-4)}{2}$)    & $1$ & $+$ \\
$n\geq3$   &$\frac{n+5}{2}$&$\frac{3}{2}$&($\frac{(n-2)}{2},\frac{(n-2)}{2},\frac{(n-2)}{2},\frac{-(n-4)}{2}$) & $2$ & $+$ \\
$n\geq4$   &$\frac{n+5}{2}$&$\half$      &($\frac{n}{2},\frac{(n-2)}{2},\frac{(n-4)}{2},\frac{-(n-4)}{2}$)    & $2$ & $+$ \\
$n\geq4$   &$\frac{n+7}{2}$&$\half$      &($\frac{(n-2)}{2},\frac{(n-4)}{2},\frac{(n-4)}{2},\frac{-(n-4)}{2}$)& $4$ & $+$ \\  
$n\geq4$   &$\frac{n+6}{2}$&$1$          &($\frac{(n-2)}{2},\frac{(n-2)}{2},\frac{(n-4)}{2},\frac{-(n-4)}{2}$)& $3$ & $+$ \\
$n\geq4$   &$\frac{n+4}{2}$&$0$          &($\frac{n}{2},\frac{n}{2},\frac{(n-4)}{2},\frac{-(n-4)}{2}$)        & $2$ & $+$ \\
$n\geq4$   &$\frac{n+6}{2}$&$0$          &($\frac{n}{2},\frac{(n-4)}{2},\frac{(n-4)}{2},\frac{-(n-4)}{2}$)    & $3$ & $+$ \\  
$n\geq4$   &$\frac{n+8}{2}$&$0$          &($\frac{(n-4)}{2},\frac{(n-4)}{2},\frac{(n-4)}{2},\frac{-(n-4)}{2}$)& $6$ & $+$ \\
        
	\hline
        
$n=1$      &   $2$         &$\half$      &($\half,\half,\half,\half$) & $1$ & $-$ \\
$n=1$      & $\frac{5}{2}$ &$0$          &($\half,\half,\half,-\half$)& $2$ & $-$ \\
        \hline 
        
$n=2$      &   $3$         &$0$          &($1,1,0,0$)                 & $2$ & $-$ \\
$n=2$      & $\frac{7}{2}$ &$\half$      &($1,0,0,0$)                 & $2$ & $-$ \\
$n=2$      & $4$           &$1$          &($0,0,0,0$)                 & $3$ & $-$ \\
        \hline
        \hline

\end{tabular}
}

It is now a simple matter to compute the Index over single gravitons.
The Witten Index for the $n^{th}$ graviton representation ($R_n$) is 
given by 
\begin{equation}\label{chip} 
\begin{split}
I_{R_n}^W &=  {\rm Tr}_{\Delta = 0} \Big[
{(-1)^F x^{\epsilon_0 + j}y_1^{H_1}y_2^{H_2}y_3^{H_3}} \Big]\\
&=\sum_{q}\frac{(-1)^{2j_q}x^{(\epsilon_0 + j)_q}\chi_q^{SO(6)}(y_1,y_2,y_3)}{1-x^2},
\end{split}
\end{equation}
where $q$ runs over all conformal representations with $\Delta=0$ that 
appear in the decomposition of $R_n$ in table \ref{tb:d3gravspec}. 
$H_1,H_2,H_3$ are the Cartan charges of $SO(6)$ in the `orthogonal'  basis 
that we always use in this paper. $\chi^{SO(6)}$, the $SO(6)$  character, 
 may be computed using the Weyl character formula. 
The full index over single gravitons is 
\begin{equation}\label{indspa} \begin{split}I_{sp} = 
\sum_{n=2}^{\infty}I^W_{R_n}+I^W_{R_1},\end{split}
\end{equation} 
After some algebra we find 
\begin{equation}\label{indspb} 
\begin{split}
I_{sp} =& \Big[ -x \left(x^2-1\right) y_1 y_2 y_3^2+\sqrt{x} \sqrt{y_1}
   \sqrt{y_2} \left(x^3-y_2+y_1 \left(x^3 y_2-1\right)\right)
   y_3^{3/2}\\&-x \left(x^2-1\right) \left(y_1+y_2\right) \left(y_1
   y_2+1\right) y_3+\sqrt{x} \sqrt{y_1} \sqrt{y_2} \left(y_2
   x^3+y_1 \left(x^3-y_2\right)-1\right)\\& \sqrt{y_3}-x
   \left(x^2-1\right) y_1 y_2 \Big]  / \Big[ \left(x^2-1\right) \left(\sqrt{x}
   \sqrt{y_1} \sqrt{y_2}-\sqrt{y_3}\right)\\& \left(\sqrt{x}
   \sqrt{y_1} \sqrt{y_3}-\sqrt{y_2}\right) \left(\sqrt{x}
    \sqrt{y_2} \sqrt{y_3}-\sqrt{y_1}\right)
   \left(\sqrt{x}-\sqrt{y_1} \sqrt{y_2} \sqrt{y_3}\right) \Big]
\end{split}
\end{equation}

The index over the Fock-space of gravitons may now be obtained from the 
above single particle index using 
\begin{equation}\label{indfs} \begin{split}I_{fock} =\exp {\left( \sum_{n}{1 \over
n}I_{sp}(x^n,y_1^n,y_2^n,y_3^n) \right)}.\end{split}
\end{equation}

In order to get a feel for this result, let us set $y_i=1$. The single 
graviton index  reduces to 
\begin{equation}\label{indsg} 
\begin{split}
I_{sp} = \frac{2 \sqrt{x} \left(2
   x+\sqrt{x}+2\right)}{\left(\sqrt{x}-1\right)^2 (x+1)}
\end{split}
\end{equation}
In the high energy limit, $x \equiv e^{-\beta}  \ra 1$,  
this expression simplifies to $I_{sp} \approx \frac{20}{\beta^2}$ 
In this limit the expression for the full Witten Index $I_{fock}$ in 
\eqref{indfs} reduces to,
\begin{equation}
I_{fock} \approx \exp{\frac{20\zeta(3)}{\beta^2}}
\end{equation}
It follows that the thermodynamic expectation value of $\epsilon_0+j$ 
(which we denote by $ E_{\rm av}^{\rm ind}$) is given by 
\begin{equation} \begin{split} E_{\rm av}^{\rm ind}  =
-{\frac{\partial \ln I_{fock}}{\partial \beta}} = 
{40\zeta(3) \over \beta^3}.\end{split}
\end{equation}
The index `entropy' defined by
\begin{equation}
I_{fock} = \int dy \exp\{(-\beta y)+ S_{\rm ind}(y)\},
\end{equation} 
evaluates to 
\begin{equation} S_{\rm ind}(E) =
\frac{60\zeta(3)}{(40\zeta(3))^{\frac{2}{3}}} E^{2 \over 3}.
\end{equation}

It is instructive to compare this result with the relation between
entropy and $E$ computed from the  
supersymmetric partition function, obtained by summing over all supersymmetric 
states with no $(-1)^F$ -- once again in the gravity approximation.
The single particle partition function evaluated on the $\Delta = 0$ 
states with all the other chemical potentials except the one corresponding to 
$E=\epsilon_0+j$ set to 
zero is given by,
\begin{equation}\label{ptfnsp} 
\begin{split}
Z_{sp}(x)= {\rm tr}_{\Delta = 0} x^{E} = \frac{2 \sqrt{x} (x+1) \left(x^{5/2}-2 x^2+2 x^{3/2}+2 x-3
   \sqrt{x}+2\right)}{\left(\sqrt{x}-1\right)^4 \left(x^2-1\right)},
\end{split}
\end{equation}
where once again $x \equiv e^{-\beta}$, with $\beta$ being the chemical potential corresponding to $E=\epsilon_0+j$.
The bosonic and fermionic contributions to the partition function 
in \eqref{ptfnsp} are respectively given by,
\begin{equation}
Z_{sp}^{\rm bose}(x)= {\rm tr}_{\Delta = 0~ {\rm bosons}} x^{E} = \frac{-\left(-x^4+4 x^{7/2}-6 x^3+x^2-4 x^{3/2}+6 x-4
   \sqrt{x}\right)}{\left(1-\sqrt{x}\right)^5 \left(\sqrt{x}+1\right)
   (x+1)}
\end{equation}
\begin{equation}
Z_{sp}^{fermi}(x)={\rm tr}_{\Delta = 0~ {\rm fermions}} x^{E} = \frac{-\left(-x^4+x^2-4 x^{3/2}\right)}{\left(1-\sqrt{x}\right)^5
   \left(\sqrt{x}+1\right) (x+1)}
\end{equation}

To obtain the index on the Fock space, we need to multi-particle the partition
function above with the correct Bose-Fermi statistics. This leads to
\begin{equation}
Z_{fock} = \exp{\sum_n {Z_{sp}^{bose}(x^n) + (-1)^{n+1}Z_{sp}^{fermi}(x^n) \over n}}. 
\end{equation}
We find, that for $\beta << 1$ 
\begin{equation}
\ln Z_{fock} = \frac{63\zeta(6)}{\beta^5},
\end{equation}
and a calculation similar to the one done above yields
\begin{equation} 
\label{m2susygravgrowth}
S(E) = \frac{378\zeta(6)}{(315\zeta(6))^{\frac{5}{6}}} E^{5 \over 6}.
\end{equation}
which is the growth of states with energy of a six dimensional gas, an 
answer that could have been predicted on qualitative grounds. Recall that 
the theory of the worldvolume of the $M2$ brane has 4 supersymmetric 
transverse fluctuations and one supersymmetric derivative. Bosonic 
supersymmetric gravitons are in one to one correspondence with `words' 
formed by acting on symmetric combinations of these scalars with an arbitrary 
number of derivatives. Consequently, supersymmetric gravitons are labelled 
by 5 integers $n_i$, $n_d$ (the number of occurrences of each of these four 
scalars $i=1 \ldots 4$ and the derivative $n_d$) and the energy of these 
gravitons is $E=\half( \sum_i n_i ) + n_d$. This is the same as the formula 
for the energy of massless photons in a five spatial dimensional rectangular
box, four of whose sides are of length two and whose remaining side is of unit
length, explaining the effective six dimensional growth.     

We conclude that the growth of states in 
the effective index entropy is slower than the growth of supersymmetric 
states in the system; this is a consequence of partial Bose-Fermi 
cancellations (due to the $(-1)^F$). 

\subsection{The Index on the worldvolume theory of a single $M2$ brane}
We will now compute our index over the worldvolume theory of a single $M2$ brane. For this free theory, the single particle state content is just the representation corresponding to $n=1$ in Table \ref{tb:d3gravspec} of the previous subsection. This means that it corresponds to the representation of the $d=3$ superconformal group with the primary having charges $\epsilon_0 = {1 \over 2}, j = 0$ and $SO(8)$ highest weights (in the convention described above) $[\half, \half, \half, -\half]$. 

For the reader's convenience, we reproduce the conformal multiplets that appear in this representation in the Table below. Physically, these multiplets correspond to the 8 transverse scalars, their fermionic superpartners and the equations of motion for each of these fields.\footnote{Please see \cite{Seiberg:1997ax,Minwalla:1998rp} and references therein for more details on this worldvolume theory and \cite{Bagger:2007vi} for some recent work.}
\begin{equation}
\begin{array}{|c|c|c|c|c|} 
\hline
{\rm letter}& \epsilon_0& j&  [h_1, h_2, h_3, h_4]  & \Delta = \epsilon_0 - j - h_1\\
\hline 
\phi^a & {1 \over 2} & 0 & [\half, \half, \half, -\half] & 0\\
\psi^a & 1 & {1 \over 2} &[\half, \half, \half, \half] & 0  \\
\pslash \psi^a = 0 & 2 & {1 \over 2} & [\half, \half, \half, \half]&1 \\
\partial^2 \phi^a = 0& {5 \over 2} & 0 & [\half, \half, \half, -\half] & 2\\ 
\hline
\end{array}
\end{equation}

The Index over these states is 
\begin{equation}
\label{m2singleindex}
\begin{split}
I_{M_2}^{\rm sp}(x,y_i) &=  {\rm Tr} \left[{(-1)^F x^{\epsilon_0 + j}y_1^{H_1}y_2^{H_2}y_3^{H_3}} \right] \\
&= {x^{1 \over 2}\left( 1+ y_1 y_2 + y_1 y_3 + y_2 y_3 \right) - x^{3 \over 2}  \left(y_1 + y_2 + y_3 + y_1 y_2 y_3\right) \over \left(y_1 y_2 y_3\right)^{1 \over 2}  \left( 1 - x^2 \right)}
\end{split}
\end{equation}

For simplicity, let us set $y_i \rightarrow 1$. Then, we find
\begin{equation}
I_{M_2}^{\rm sp} (x,y_i=1) = {4 x^{1 \over 2} \over 1 + x}
\end{equation}
Multiparticling this index, to get the index over the Fock space on the $M_2$ brane, we find that 
\begin{equation}
\label{indexmultipart}
\begin{split}
I_{M_2}(x, y_i = 1) &= \exp{\sum_{n \geq 1} {I_{M_2}(x^n,y_i = 1) \over n}}\\
&= \left(\prod_{n \geq 0} {1 - x^{2 n + {3 \over 2}} \over 1 - x^{2 n + {1 \over 2}}} \right)^4
\end{split}
\end{equation}
At high temperatures $x \equiv e^{-\beta} \rightarrow 1$, the index grows as
\begin{equation}
\left. I_{M_2} \right|_{x \rightarrow 1, y_i = 1} = \left({\beta \over 2} \right)^{-2}
\end{equation}

The single particle supersymmetric partition function, obtained by summing over  all $\Delta = 0$ single particle states with no $(-1)^F$ is,
\begin{equation}
\begin{split}
Z^{\rm susy, sp}_{M_2}(x, y_i) &= {\rm Tr}_{\Delta = 0} \left[{x^{\epsilon_0 + j}y_1^{H_1}y_2^{H_2}y_3^{H_3}} \right] \\
&={x^{1 \over 2}\left( 1+ y_1 y_2 + y_1 y_3 + y_2 y_3 \right) + x^{3 \over 2}  \left(y_1 + y_2 + y_3 + y_1 y_2 y_3\right) \over \left(y_1 y_2 y_3\right)^{1 \over 2}  \left( 1 - x^2 \right)}
\end{split}
\end{equation}
Setting $y_i \rightarrow 1$, 
\begin{equation}
Z_{M_2}^{\rm susy, sp} (x,y_i=1) = {4 x^{1 \over 2} \over 1 - x}
\end{equation}
with individual contributions from bosons and fermions being
\begin{equation}
\begin{split}
Z_{M_2}^{\rm susy,sp, bose}(x) &= {\rm tr}_{\Delta=0~{\rm bosons}} x^{E} = {4 x^{1 \over 2} \over (1 - x^2)} \\
Z_{M_2}^{\rm susy,sp, fermi}(x) &= {\rm tr}_{\Delta=0~{\rm fermions}} x^{E} = {4 x^{3 \over 2} \over (1 - x^2)} 
\end{split}
\end{equation}
Finally, multi-particling this partition function with the appropriate bose-fermi statistics, we find that
\begin{equation}
\label{susymultipart}
Z_{M_2}(x, y_i = 1) = \left(\prod_{n \geq 0} {1 + x^{2 n + {3 \over 2}}  \over 1 - x^{2 n + {1 \over 2}}} \right)^4
\end{equation}

At high temperatures $x \rightarrow 1$, the supersymmetric partition function grows as
\begin{equation}
Z_{M_2}(x \rightarrow 1, y_i = 1) \approx \exp\left\{{\pi^2 \over 2 \beta}\right\}
\end{equation}
Note, that this partition function grows significantly faster at high temperatures than the index \eqref{indexmultipart}. 

\subsection{Index over Chern Simons Matter Theories}

In this subsection, we will calculate the Witten Index described above 
for a class of the superconformal Chern Simons matter theories recently 
studied 
by Gaiotto and Yin \cite{Gaiotto:2007qi}. The theories studied by these authors 
are three dimensional Chern Simons gauge theories coupled to matter fields; 
we will focus on examples that enjoy invariance under a superalgebra 
consisting of $4$ Qs and $4$ Ss (i.e. the $R$ symmetry of these theories  
is $SO(2)$). The matter fields, which  may thought of as dimensionally reduced 
$d=4$ chiral multiplets, carry the only propagating degrees of freedom.
The general constructions of Gaiotto and Yin allow the possibility of nonzero 
superpotentials with a coupling $\alpha$ that flows in the infra-red to a fixed point of order ${1 \over k}$ where $k$ is the level of the Chern Simons theory.
In our analysis below we will focus on the limit of large $k$. In this limit,
the theory is `free' and moreover we may treat ${1 \over k}$ as a
 continuous parameter. The arguments above then indicate index that we compute below for the free theory will be invariant under small deformations
 of ${1 \over k}$.

Consider this free conformal 3 dimensional theory on $S^2$. We 
are interested in calculating the letter partition function (i.e. the 
single particle partition function) for the propagating fields which comprise
a complex scalar $\phi$ and its fermionic superpartner $\psi$. This may be 
done by enumerating all operators, 
linear in these fields, modulo those operators that are set to 
zero by the equations of motion. We will be interested in keeping track of 
several charges: the energy $\epsilon_0$, $SO(3)$ angular momentum $j$,  
$SO(2)$ R-charge $h$ and $\Delta=\epsilon_0 - h - j$ of our states. 
The following table (which lists these charges) is useful for that purpose 
\begin{equation}
\begin{array}{|c|c|c|c|c|} 
\hline
{\rm letter}& \epsilon_0& j&  h  & \Delta = \epsilon_0 - j - h\\
\hline 
\phi & {1 \over 2} & 0 & {1 \over 2} & 0\\
\phi* & {1 \over 2} & 0 & {-1 \over 2} & 1\\
\psi & 1 & {1 \over 2} &{-1 \over 2} & 1  \\
\psi* & 1 & {1 \over 2} & {1 \over 2}&0 \\
\partial_{\mu} & 1 & \{\pm 1, 0\} & 0 & \{0,2,1\} \\
\partial_{\mu} \sigma^{\mu} \psi= 0 & 2 & {1 \over 2} & {-1 \over 2}&2 \\
\partial_{\mu} \sigma^{\mu} \psi^* = 0 & 2 & {1 \over 2} & {1 \over 2}&1 \\
\partial^2 \phi = 0& {5 \over 2} & 0 & {1 \over 2} & 2\\ 
\partial^2 \phi^* = 0 & {5 \over 2} & 0 & {-1 \over 2}&3\\
\hline
\end{array}
\end{equation}
The last four lines, with equations of motion count with minus signs in the partition function. The list above comprises two separate irreducible representations of the superconformal algebra. $\phi$, $\psi$ and derivatives on these letters make up one representation. The other representation consists of the conjugate fields.

Let the partition functions over these two representations be denoted by 
$z_1$ and $z_2$. We find 
\begin{equation}
\begin{split}
z_1[x,y,t] &= {\rm tr}_{\phi, \psi, \ldots} (x^{2 \epsilon_0} y^{2 j} t^{h}) = {t^{1 \over 2}  x(1 + x^2)+  t^{-1 \over 2} x^2 (y + 1/y) \over (1 - x^2 y^2) (1 - x^2/y^2)} \\
z_2[x,y,t] &= {\rm tr}_{\phi^*, \psi^*, \ldots} (x^{2 \epsilon_0} y^{2 j} t^{2 h}) = {t^{-1 \over 2} x(1 + x^2)+  t^{1 \over 2} x^2 (y + 1/y) \over (1 - x^2 y^2) (1 - x^2/y^2)}
\end{split}
\end{equation}
The index \eqref{wia} over single particle states is obtained by setting  
$t \rightarrow 1/x , y \rightarrow -1$
\begin{equation}
\begin{split}
I_1[x] &= z_1[x,-1 ,1/x] = {\rm tr} ((-1)^F (x)^{2 \epsilon_0 - h}) = {x^{1 \over 2} \over 1- x^2} \\
I_2[x] &= z_2[x, -1, 1/x] = {\rm tr} ((-1)^F x^{2 \epsilon_0 - h}) =  {-x^{3 \over 2} \over 1 - x^2} \\
I[x] &= I_1[x] + I_2[x] = {x^{1 \over 2} \over 1 + x}
\end{split}
\end{equation}
In terms of these quantities, the index of the full theory is given by\cite{Sundborg:1999ue,Aharony:2003sx}
\begin{equation}\label{indexthree}
I^W= \int DU \exp \left[ \sum_{n=1}^\infty \sum_m {I(x^n) \over n} Tr_{R_m} (U^n) 
\right]
\end{equation}  
where $m$ run over the chiral multiplets of the theory, which are taken 
to transform in the $R_m$ representation of $U(N)$, and $Tr_{R_m}$ is 
the trace of the group element in  the $R_m^{th}$ representation of $U(N)$.

In the large $N$ limit the  integral over $U$ in \eqref{indexthree} may 
be converted into an integral over the eigenvalue distribution of $U$,   
$\rho(\theta)$, which, in turn, may be computed via saddle points.\footnote{ Note that $N \rho(\theta) d \theta$ gives the number of 
eigenvalues between $e^{i \theta}$ and $e^{i (\theta + d \theta)}$ and 
$\int_{-\pi}^{\pi} \rho(\theta) d \theta = 1,~~\rho(\theta) \geq 0$}
The Fourier coefficients of this eigenvalue density function are given by:
\begin{equation}
\rho_n =  \int_{-\pi}^{\pi} \rho(\theta) \cos(n \theta)
\end{equation}

\subsubsection{Adjoint Matter}

In order to get a feel for this formula, we specialize to a particular 
choice of matter field content. We consider a theory with $c$ matter fields
all in the adjoint representation. In the large $N$ limit the Index is given 
by 
\begin{equation}
\label{indexpression}
\begin{split}
{\cal I}(x) &= Tr_{\rm colour singlets} (-1)^F x^{2 \epsilon_0 - h} \\
&= \int d \rho_n \exp \left(-N^2 \sum_{n=1}^{\infty} {1 \over n} (1 - c I[x^n]) \rho_n^2 \right)
\end{split}
\end{equation}
The behaviour of this Index as a function of $x$ is dramatically different for 
$c\leq 2$ and $c \geq 3$. In order to see this note that at any given 
value of $x$, the saddle point occurs at $\rho(\theta) = {1 \over 2 \pi}$ 
i.e $\rho_0 = 1, \rho_n = 0, n > 0$ provided that\cite{Sundborg:1999ue,Aharony:2003sx}
\begin{equation}\label{condtriv}
1 - c I[x^n] > 0, \forall n
\end{equation}
In this case the saddle point contribution to the Index vanishes; the 
leading contribution to the integral is then from the Gaussian 
fluctuations about this saddle point. Under these conditions the 
logarithm of the Index or the 'free-energy' 
\begin{footnote}{We use this term somewhat loosely, since we are 
referring here to an index and not a partition function}\end{footnote} 
is then of order $1$ in the $\frac{1}{N}$ expansion. 

It is easy to check that \eqref{condtriv} is satisfied at all values of $x$
(which must lie between zero and one in order for \eqref{wia} to be well 
defined) when $c\leq 2$.  On the other hand, if $c \geq 3$ this condition is 
only met for 
\begin{equation}
\label{hagtemp}
x < \left({1 \over 2} \left(c - \sqrt{c^2 - 4}\right) \right)^2
\end{equation}
At this value of $x$ the coefficient of $\rho_1^2$ in \eqref{indexpression} 
switches sign and the saddle point above with a uniform eigenvalue 
distribution is no longer valid. The new saddle point that dominates this 
integral above this value of $x$, has a Gross-Witten type gap in the 
eigenvalue distribution. The Index undergoes a large $N$  first order 
phase  transition at the critical temperature listed in \eqref{hagtemp}. 
At and above this temperature the 'free-energy' is of order $N^2$.

Note that $I(1)=\half$. It follows that the Index is well defined even 
at strictly infinite temperature  
This is unlike the logarithm of the actual partition function of 
the same theory, 
whose $x \rightarrow 1$ limit scales like $N^2/(1-x)^2$ as $x \to 1$ 
(for all values of $c$) reflecting the $T^2$ dependence of a 2+1 
dimensional field theory. 
This difference between the high temperature limits of the Index and the 
partition function reflects the large cancellations of supersymmetric 
states in their contribution to the Index. 

\subsubsection{Fundamental Matter}

As another special example, let us consider a theory whose $N_f$ matter 
fields all transform in the fundamental representation of $U(N)$. 
We take the  Veneziano  limit: 
$N_c \rightarrow \infty, c = {N_f \over N_c}$ fixed.  The index for 
the theory is now given by
\begin{equation}
\label{indexpressionfund}
\begin{split}
{\cal I}(x) &= Tr_{\rm colour singlets} (-1)^F x^{2 \epsilon_0 - h} \\
&= \int d \rho_n \exp(-N^2 \sum_{n=1}^{\infty} { (\rho_n  - c I[x^n])^2 - c^2 I[x^n]^2 \over n})
\end{split}
\end{equation}
At low temperatures the integral in \eqref{indexpressionfund} is dominated by the saddle point 
\begin{equation}
\label{saddle}
\rho_n=c I(x^n)
\end{equation}
As the temperature is raised the 
integral in \eqref{indexpressionfund} undergoes a Gross-Witten 
type phase transition when $c$ is large enough. This is easiest to 
appreciate in the limit $c \gg 1$. In this limit $\rho_1 =\half$ in the 
low temperature phase when at $x \approx \frac{1}{4 c^2}$, and 
$\rho_n=\frac{1}{2^n c^{n-1}} \ll 1$. At approximately this value of $x$ the 
low temperature eigenvalue distribution $\rho(\theta)$ formally turns 
negative at $\theta=\pi$. This is physically unacceptable (as an eigenvalue 
density is, by definition, intrinsically positive). In actual fact the 
system undergoes a phase transition at this value of $x$. At large $c$ this phase transition is very
similar to the one described by Gross and Witen in \cite{Gross:1980he} and in a more closely related context by
\cite{Schnitzer:2006xz}. The high temperature eigenvalue distribution is `gapped' i.e. it has support on only a subset
(centered about zero) of the interval $(-\pi, \pi)$.

For this phase transition to occur, we need $c \geq 3$. To arrive at this 
result, we notice that the distribution \eqref{saddle} implies
\begin{equation}
\lim_{x \rightarrow 1-} \rho(\pi)=\lim_{x \rightarrow 1-}\rho(-\pi) = {1 \over \pi} \left({1 \over 2} - {c \over 4} \right)
\end{equation}
So, for $c \geq 3$, $\rho(\pi)$ would always turn negative for some value of $x$. Beyond this temperature the saddle point \eqref{saddle} is no longer valid.

\section{d=6}\label{dsamansix}

\subsection{The Superconformal Algebra and its Unitary Representations}\label{subsec:dsixintro}
The bosonic subalgebra of the $d=6$ superconformal algebra is  
$SO(6,2) \otimes Sp(2n)$ (the conformal algebra times the R symmetry algebra). 
The anticommuting generators in this algebra may be divided into 
the generators of supersymmetry ($Q$) and the generators of 
superconformal symmetries ($S$). Supersymmetry generators transform in the 
fundamental representation of the R-symmetry group $Sp(2n),$ \footnote{With 
our conventions, $Sp(2n)$ is of rank $n$. $Sp(2)=SO(3)$ and $Sp(4)=SO(5)$.} 
have charge half 
under dilatations (the $SO(2)$ factor of the compact $SO(6)\otimes SO(2)
\in SO(6,2)$) and are chiral spinors under the $SO(6)$ factor of the same 
decomposition. Superconformal generators $S_i^{\mu} = (Q^i_{\mu})^{\dagger}$ 
transform in the anti-chiral spinor representation of $SO(6)$, have scaling 
dimension (dilatation charge) $(-\half)$, and also 
transform in the anti-fundamental 
representation of the R-symmetry group. The charges of these generators are given in more detail in Appendix \ref{appcharges}. In our
notation for supersymmetry generators $i$ is an $SO(6)$ spinor index 
while $\mu$ is an $R$ symmetry vector index.

The commutation relations for this superalgebra are described in detail
in \cite{Minwalla:1997ka}. 
As usual, the anticommutator between two supersymmetries 
is proportional to momentum times an $R$ symmetry delta function, and the 
anticommutator between two superconformal generators is obtained by 
taking the Hermitian conjugate of these relations. The most interesting 
relationship in the algebra is the anticommutator between $Q$ and $S$. 
Schematically
\begin{equation*}\{S_i^{\mu},Q^j_{\nu}\} \sim \delta^{\mu}_{\nu}T^{j}_{i} -
\delta^{j}_{i} M^{\mu}_{\nu}\end{equation*} Here $T^{ij}$ are the $U(4) \sim
SO(6)\times SO(2)$ generators and $M_{\mu\nu}$ are the $Sp(2n)$
generators. The energy $\epsilon_0$, $SO(6)$ highest weight 
( denoted by $h_1, h_2$ and  $ h_3$ in the orthogonal basis
\footnote{$h_i$ are eigenvalues under rotations in orthogonal 2 planes 
in $R^n$. Thus, for instance, $\{h_i \}=(1,0,0)$ 
in the vector representation. They are  either integer or half integer and satisfy the constraint $h_1 \geq h_2 \geq |h_3| \geq 0$})  and the R-symmetry 
highest weights ($ k, k_1 \ldots,  k_{(n-1)})$ 
of primary states form a complete set of labels for the representation 
in question. We use a non-standard normalization for the R-symmetry weights. In particular, 
\begin{equation}
k = {k^o \over 2}, ~~~ k_i = {k_i^o \over 2}
\end{equation}
Here $[k^o, k_i^o]$ are the highest weights of $Sp(2 n)$ in the orthogonal basis.\footnote{In the orthogonal basis, the Cartans of $Sp(2 n)$ are $2n \times 2n$ matrices with elements ${\rm diag}(i \sigma_2, 0, 0 \ldots ), {\rm diag}(0, i \sigma_2, 0, 0, \ldots), \ldots$, where each $0$ is shorthand for a $2 \times 2$ matrix}
As we have noted above, at the level of the algebra, 
$SO(2) \times SO(6) \sim U(4)$. We will sometimes find it convenient to label 
primaries by eigenvalues $c_i$ under the generators $T^i_i \equiv T_i$
of $U(4)$\footnote{In the defining representation of $U(4)$ $(T_i)^a_b =
\delta^a_i \delta_{b}^{i}.$ } rather than by the energy and $SO(6)$ weights. 
For any highest weight $(c_1,c_2,c_3,c_4)$ the eigenvalues satisfy  $c_1 \geq c_2 \geq c_3 \geq c_4 \geq 0$ and 
$c_i$ s are always integers. For future reference we note the change of 
basis between the Cartan elements  $\epsilon_0, h_1, h_2, h_3$ (the energy and 3 
orthogonal $SO(6)$ Cartan generators) and 
$T_1, T_2, T_3, T_4$: 
\begin{equation}\label{tran}
 \begin{split}{\epsilon_0 = &\  \half(T_1 + T_2 + T_3 + T_4)\\
                    h_1 = &\ \half(T_1 + T_2 - T_3 - T_4)\\
                    h_2 = &\ \half(T_1 - T_2 + T_3 - T_4)\\
                    h_3 = &\ \half(T_1 - T_2 - T_3 +
                    T_4)\\}\end{split}
                    \end{equation}

As in the case of the $d=3$ algebra, any irreducible representation of 
the superconformal algebra may be decomposed into a finite number of 
distinct irreducible representations  of the conformal algebra. The latter 
are labeled by their own conformal primary states, which have a definite 
lowest energy and transform in a given irreducible representation of $SO(6)$. 

We will now analyse the constraints imposed by unitarity on the quantum 
numbers of primary states; for this purpose we will find it convenient 
to use the $U(4)$ labeling of primaries introduced above. Let $Q^i_{\m}\ i = 1,
\cdots, 4.\ $ and $\m = \pm 1, \cdots, \pm n$ denote the
supersymmetry whose charge under $U(4)$ Cartan $T_j$ are $\delta^i_j$
and  under the R-symmetry Cartan $M_{\n}$ is ${\rm (sign\ of \m) }\times
\delta^{\n}_{|\m|}$. The superconformal generators are 
$S_i^{\m} = (Q^i_{\m})^{\dagger}$ and therefore they have the same 
charges as $Q^i_{\m}$ but with opposite sign. 

\subsection{Norms and Null States}
\label{nullstructure6d}
In this subsection we study unitarity restrictions (and the resultant 
structure of null states) of representations of the superconformal 
algebra. This analysis turns out to be a little more intricate than 
its $d=3$ counterpart. 

As we have seen above, states in the same representation of the superconformal 
algebra do not all have the same norm. However states that lie within the 
same representation of the maximal compact subgroup of the algebra, 
$U(4) \times Sp(2n)$, do have the same norm. Consequently, in order to 
examine the constraints from unitarity, we need only examine one state 
per representation of this compact subalgebra. 

In order to study the restrictions imposed by unitarity at level $\ell$ we should,
in principle, study all states obtained by acting with the tensor product 
of an arbitrary combination of $\ell$ supersymmetries on the set of primary 
states of an irreducible representation of the superconformal algebra. 
This set of states may be Clebsh Gordan decomposed into a sum of irreducible 
representations of $U(4) \times Sp(2n)$; and we should 
compute the norm of at least one state in each of these representations, and 
ensure its positivity in order to guarantee unitarity. However
this problem is significantly simplified by the observation that 
the most stringent condition on unitarity occurs in those states that 
transform in the `largest' $Sp(2n)$ \cite{Dobrev:2002dt}. Now it is easy to construct a state in the 
largest $Sp(2 n)$ representation: one simply acts on those primary states that 
are $Sp(2n)$ highest weight with $\ell$ $Sp(2n)$ highest weight supersymmetries, 
i.e. supersymmetries of the form $Q^i_1$. This prescription completely fixes 
the $Sp(2n)$ quantum numbers of the states we will study in this section. 
All that remains is to study the decomposition of all such states into 
irreducible representations of $U(4)$ and to compute the norm of one state
in each of these representations. 

The decomposition of the states of interest into $U(4)$ representations at 
level $\ell$ is easily performed using Young Tableaux techniques. The set of 
U(4) tableaux for representations of the descendants is obtained by 
adding $\ell$ boxes to the tableaux of the primary in all possible ways that 
give rise to a legal tableaux, subject to the restriction that no two 
`new' boxes occur on the same row (this restriction is forced on us by 
the antisymmetry of the $Q_1^i$ operators). Note, that in this decomposition,
no representation occurs more than once.\footnote{For a generic primary tableaux 
the number of representations obtained at level $\ell$ is ${4 \choose \ell}$ 
corresponding to the choice of which rows the new boxes are appended to.
If the $U(4)$ highest weights of the primary are $c_1, c_2, c_3, c_4$, 
the representation obtained by appending new boxes to the rows $R^{i_1}$, 
$R^{i_\ell}$ has highest weights $c_{i_1} ... c_{i_\ell}$ increased by one, 
while all other weights are unchanged.} 

It is not too difficult to find an explicit formula for the highest weight 
states of each of these representations.  Let us define the operators 
$\left(A^i= \sum_{j =1}^iQ_1^j\Upsilon_j^i
\right)\ \ i = 1,\cdots, 4$ where $\Upsilon_j^i$ are functions of the 
$U(4)$ generators defined by   
\begin{equation}\label{upa}
\begin{split}\Upsilon_j^j =& {\rm Identity\ \ (no\ sum\ over\ j)}\\
                  \Upsilon_1^4 =& -\Big[ T^2_1 T^3_2 T^4_3 \left({\frac{(T_3-T_4 +1)(T_2-T_4+2)}{(T_3-T_4)(T_2 -T_4
	 	+1)}}\right) \\ &- T^3_2 T^4_3 T^2_1\left({\frac{T_3 - T_4 +1}{T_3 - T_4}} \right) - T^4_3 T^2_1 T^3_2
	  	\left({\frac{T_2 -T_4 +2}{T_2 -T_4 +1}} \right) 
                  + T^4_3 T^3_2 T^2_1 \Big] \left({\frac{1}{T_1 -T_4 +2}}\right)\\
                   \Upsilon_2^4 =& -\left(T^4_3T^3_2 - T^3_2T^4_3
		 \left({\frac{T_3-T_4+1}{T_3-T_4}}\right)\right)\left(\frac{1}{T_2 -T_4 +1}\right)\\
                   \Upsilon_3^4 =& -T^4_3\left({\frac{1}{T_3 -T_4}}\right)\\
                   \Upsilon_1^3 =& -\left(T^3_2T^2_1 - T^2_1T^3_2\left({\frac{T_2 -T_3
		 +1}{T_2-T_3}}\right)\right)\left({\frac{1}{T_1-T_3+1}}\right)\\
                   \Upsilon_2^3 =& -T^3_2\left({\frac{1}{T_2-T_3}}\right)\\
                   \Upsilon_1^2 =& -T^2_1\left({\frac{1}{T_1 - T_2}}\right)
\end{split}
\end{equation}
The operators $A^i$ have been determined to have the following property: 
when acting on a highest weight state $|\psi\rangle$ of $U(4)$ with quantum 
numbers $(c_1, c_2, c_3, c_4)$, $A^i|\psi \rangle$  
is another highest weight state of $U(4)$ with quantum numbers 
$(c^i_1, c^i_2, c^i_3, c^i_4)$ where $c^i_j=c_j + \delta_i^j$, whenever it is 
well defined. The last condition (being well defined) is met if and only 
if the weights of $|\psi \rangle$ obey the inequality $c_i < c_{i- 1}$
\footnote{This is rather intuitive; when this condition is not met, the set 
$(c^i_1, c^i_2, c^i_3, c^i_4)$ do not constitute a valid set of labels for 
an irreducible representation of $U(4)$.} 

Let $|\psi \rangle$ denote the primary state that is a $U(4)$ highest weight. 
It follows that the states $A^{i_1}... A^{i_\ell} |\psi 
\rangle$ is the highest weight state in the representation with additional 
boxes in the rows $i_1... i_\ell$ described above. We will now study the norm 
of these states. 

It is not difficult to explicitly verify that (when this state 
is well defined) 
\begin{equation} \label{levelonenorm}
| A^i |\psi\rangle |^2 \propto (c_i -2k -i+1) \equiv B_i(c_i, k)
\end{equation} 
More generally, it is also true that 
\begin{equation} \label{levelgennorm}
|\prod_{m=1}^l A^{i_m} |\psi\rangle |^2 \propto \prod_{m=1}^l 
B_i(c_{i_m}, k)
\end{equation} 
where the proportionality factor in \eqref{levelgennorm} is a function 
of the the $SU(3)$ weights $c_i-c_j$ of the representation but is independent 
of the energy.\footnote{More precisely, the proportionality factor is a 
function of the $c_i$ that is invariant under a uniform constant shift of 
each $c_i$.} In order to see this note that different states of the form 
\eqref{levelgennorm}, obtained by interchanging the order of the $A^{i_m}$ 
operators, are each proportional to the highest weight state of a given 
representation. Now no $U(4)$  representation occurs more than once in the tensor product of supersymmetry generators with the primary, these representations
are proportional to each other. As the commutator 
of $A^i$ operators is independent of energy, it follows that 
the proportionality factor between these states is also independent of energy. 

Now the norm of the state in \eqref{levelgennorm} clearly has a factor of 
$B_{i_l}(c_{i})$ in it. However upon interchanging the order of the $A^i$ 
factors, the same result is true for $B_{i_m}$ for each of $m=1$ to $l$. 
The norm of a state at level $\ell$ is 
a polynomial of degree $\ell$ in the energy of the state. It follows that 
the full energy dependence of the norm of this state is given as in 
\eqref{levelgennorm}; the proportionality factor in that equation is a function
only of $SU(3)$ weights and is independent of energy. 

The proof presented above, strictly speaking, applies only when each of the 
operators $A^{i_m}$ has well defined action on $|\psi \rangle$. However, 
as the algebra involved in computing \eqref{levelgennorm} is smooth 
(it does not care about the values of $c_i$ provided only that the state 
on the LHS of \eqref{levelgennorm} is well defined), and so the result 
\eqref{levelgennorm} continues to apply, whenever the state whose norm is 
being computed is well defined. 

The unitarity restrictions and short 
representations of this superconformal algebra now follow almost immediately
from \eqref{levelgennorm}. First consider the generic case representation 
where $(c_1 > c_2 > 
c_3 > c_4)$. All states listed in \eqref{levelgennorm} are well defined 
in this case and it follows $c_4 - 3 - 2k \geq 0$ is necessary and sufficient 
for unitarity. Representations that saturate this bound are short; the 
zero norm primary state is 
\begin{equation}\label{zero} |Z_4\rangle = A^4 |h.w\rangle
\end{equation}
consistent with the result of \cite{Kinney:2005ej}. 

The state \eqref{levelgennorm} is not well defined when $c_3=c_4$. However 
even in this case the state $\left(A^4 A^3\right) |\psi \rangle$ is well defined 
provided $c_2 \neq c_3$.  The norm of this state is proportional to 
$B_4 \times B_3$. A little thought shows that the necessary and sufficient
condition for unitarity is either $B_4 \geq 0$ (this is \eqref{zero}) 
or that $B_3=0$. In the later case the representation is short, and its 
level one zero norm primary is $A^3 |\psi \rangle$. On the other hand 
when $B_4=0$ the representation is also short. It's zero norm primary 
occurs at level 2 and is $\left(A^4 A^3\right) |\psi \rangle$. 

It is clear that this pattern generalizes simply. If $c_4=c_3=c_2$ but 
$c_2 \neq c_1$ then the necessary and sufficient condition for unitarity
is either $B_4 \geq 0$ or $B_3=0$ or $B_2=0$. When $B_2=0$ the zero norm 
primary occurs at level one and is given by $A^2 |\psi\rangle$. When 
$B_3=0$ the zero norm primary occurs at level 2 and is given by 
$\left(A^3 A^2\right) |\psi\rangle$. When $B_4=0$ the zero norm primary occurs at level 
3 and is given by $\left(A^4 A^3 A^2\right) |\psi \rangle $. 

Finally when $c_4=c_3=c_2=c_1$ the necessary and sufficient condition for 
unitarity is either $B_4\geq0$ or $B_3=0$ or $B_2=0$ or $B_1=0$. 
When $B_1=0$ the level one primary is given by $A^1 |\psi \rangle$. 
When $B_2=0$ the level two primary is given by $\left(A^2 A^1\right) |\psi \rangle$. 
When $B_3=0$ the level three primary is given by $\left(A^3 A^2 A^1\right) | \psi \rangle$.
When $B_4=0$ the level four primary is given by 
$\left(A^4 A^3 A^2 A^1 \right)|\psi \rangle$.

We may translate the analysis of zero norm states above into $SO(2) \times SO(6)$ notation by using the transformations of \eqref{tran}. This yields the result
that representations are short if the energy $\epsilon_0$ and $SO(6)$ weights $h_i$ satisfy one of the following conditions (see \cite{Minwalla:1997ka,Dobrev:2002dt})

\begin{equation}
\begin{split}
\epsilon_0 =& h_1 + h_2 -h_3 + 4k + 6,\ \ \rm{when} \ h_1 \geq h_2 \geq |h_3|.\\
\epsilon_0 =& h_1 + 4k + 4,\ \ \rm{when} \  h_1 \geq h_2 \  \rm{and} \   h_2 = h_3.\\
\epsilon_0 =& h_1 + 4k + 2,\ \ \rm{when} \ h_1 = h_2 = h_3 \neq 0.\\
\epsilon_0 =& 4k,\ \ \rm{when}\ h_1 = h_2 = h_3 = 0. 
\end{split}
\end{equation}
The last three conditions give isolated short representations.

\subsection{Null Vectors and Character Decomposition of a Long Representation at the Unitarity Threshold}

As discussed in the previous subsection, just like $d=3$ the short 
representations
of $d=6$ super-conformal algebra can be broadly classified into two types, the 
{\it regular} ones and the {\it isolated} ones. However unlike $d=3$ here the 
isolated short representations are of three kinds as we describe below. 
The energy of a regular short representations is given by 
$\epsilon_0 =h_1+h_2-h_3+4k+6$.
The null states of this representation also transform in an irreducible representation of the 
algebra; for $h_1>h_2$ and $h_2-\half > |h_3-\half|$ the 
highest weights of the 
primary at the head of this (null) irreducible representation (which occurs at level 1) are given 
in terms of the highest weight of the representation by 
$\epsilon_0' = \epsilon_0 +
\half,~~~h_1'=h_1-\half,~~~h_2'=h_2-\half,~~~h_3'=h_3+\half,
~~~k'=k+\half,~~~k_i'
=k_i$ (where $i=1,2,....,(n-1)$) and $k,k_i$ are half the weights of the R-symmetry group  $Sp(2n)$
in the orthogonal basis
 as defined in subsection (\S\S \ref{subsec:dsixintro}).
Note that $\epsilon_0'-h_1'-h_2'+h_3'-4k'-6 = 
\epsilon_0-h_1-h_2+h_3-4k-6
=0$, so that the null states also transform in a regular 
short representation. As 
union of the ordinary and null state of such short representations 
is identical 
to the state content of a long representation at 
the edge of the unitarity bound, 
we conclude that,

\begin{equation}\label{cenl}
\begin{split}
\lim_{\delta \ra
0}\ch[h_1+&h_2-h_3+4k+6+\delta, h_1, h_2, h_3, k, k_i]\\ &= \ch[h_1+h_2-h_3+4k+6, h_1, h_2, h_3, k, k_i]
\\&+ \ch[h_1+h_2-h_3+4k+\frac{13}{2},h_1-\half,h_2-\half,h_3+\half,k+\half,k_i],\\
&~~~~~~~~~~~~~~~~~~~~~~~~~~~~~~~~~~~~~~~~~~(\rm{with} \  h_1 > h_2 > |h_3| \geq 0).
\end{split}
\end{equation}

where $\chi(\epsilon_0,h_1,h_2,h_3,k,k_i)$ denotes the character of the irreducible representation 
of super-conformal algebra with energy $\epsilon_0$, $SO(6)$ highest weight 
$(h_1,h_2,h_3)$ and $Sp(2 n)$ highest weight $(k, k_i)$.

On the other hand, when $h_1>h_2=h_3(=h$, say$)$ the null states of the regular 
short representation occur at level 2 and are labelled by a primary with
highest weights  $\epsilon_0'=\epsilon_0+1,~~~h_1'=h_1-1,~~~h_2'=h_2=h,~~~h_3'=
h_3=h,~~~k'=k+1,~~~k_i'=k_i$, where $\epsilon_0, h_i,k,k_i$ refer to the highest 
weights of the original representation.
Note in particular that $h_2'=h_3'$ and 
$\epsilon_0'-h_1'-4k'-4 = \epsilon_0-h_1-h_2+h_3-4k-6=0$. It follows that the 
null states of this representation transforms in an isolated short 
representation and we conclude,
\begin{equation}\label{cenm} 
\begin{split}
\lim_{\delta \ra
0}\ch[h_1+4k+6+\delta,h_1,h,h,k,k_i] =& \ch[h_1+4k+6,h_1,h,h,k,k_i]\\
&+ \ch[h_1+4k+7,h_1-1,h,h,k+1,k_i]\\
&~~~~~~~~~~~~~~~~(\rm{with} \  h_1 > h_2 = h_3 = h \geq 0).
\end{split}
\end{equation}

As we have discussed earlier isolated short representations are separated from 
all other representations with the same $SO(6)$ and $Sp(2 n)$ 
quantum numbers by a 
gap in energy. Hence it is not possible to {\it approach} such a 
representation 
with long representations; consequently we have no equivalent of \eqref{cenm} 
at energies equal to $h_1+4k+7+\delta$.

Similarly when $h_1=h_2=h_3(=h \neq 0)$ the null states of the regular 
representation occur at level 3 and are labelled by a primary with highest 
weights $\epsilon_0'=\epsilon_0+\frac{3}{2},~~~h_1'=h-\half,~~~h_2'=h-\half,~~~
h_3'=h-\half,~~~k'=k+\frac{3}{2}$. Note in particular that $h_1'=h_2'=h_3'$ and 
$\epsilon_0'-h_1'-4k'-2=\epsilon_0-h_1-4k-6=0$. Consequently the null states 
of this representation transforms in an isolated short representation, and we 
conclude,

\begin{equation}\label{cenn}
\begin{split}
\lim_{\delta \ra
0}\ch[h+4k+6+\delta,h,h,h,k,k_i] =&\ch[h+4k+6,h,h,h,k,k_i]\\
&+ \ch[h+4k+\frac{15}{2},h-\half,h-\half,h-\half,k+{3\over 2},k_i].\\
&~~~~~~~~~~~~~~~~(\rm{with} \  h_1 = h_2 = h_3 = h > 0)
\end{split}
\end{equation}
As explained above, we have no equivalent of \eqref{cenn} at 
energies equal to $h+4k+\frac{15}{2}+\delta$ which corresponds to the 
unitarity bound for an isolated short representation.

Finally when $h_1=h_2=h_3=0$ the null states of the regular representation 
occur at level 4 and are labelled by primary with highest weights $\epsilon_0'=
\epsilon_0+2,~~~h_1'=h_1=0,~~~h_2'=h_2=0,~~~h_3'=h_3=0,~~~k'=k+2,~~~k_i'=k_i$. 
Note in particular that in this case $h_1'=h_2'=h_3'=0$ and $\epsilon_0'-4k'=
\epsilon_0-4k-6=0$. Therefore the null states of this representation transform 
in an isolated short representation and we conclude,
\begin{equation}\label{ceno} 
\begin{split}
\lim_{\delta \ra 0}\ch[4k+6+\delta,0,0,0,k,k_i] = \ch[4k+6,0,0,0,k,k-i] +&
\ch[4k+8,0,0,0,k+2,k_i].\\
&(\rm{with} \  h_1 = h_2 = h_3 = 0 )
\end{split}
\end{equation}
There is no equivalent of \eqref{ceno} at energies equal to $4k+6+\delta$. 

As in the previous section, the analysis of the character formulae above and 
the definition of indices is much simplified by the introduction of some additional notation. Given a short representation we will use the notation $c(h_1, h_2, h_3, k, k_i)$ to refer to this representation where the relationship between the numbers $h_i, k, k_i$ and the highest weights of the representation in question is defined in Table \ref{dsixnota}. 
\TABLE{
\caption{Notations for short representations}\label{dsixnota}
\begin{tabular}{|c|c|c|c|c|}
	\hline 
notation for rep.&$\epsilon_0$&$SO(6)$&$Sp(2 n)$&nature\\
&&highest&highest&of rep\\
&&weight&weight&\\

        \hline

$c(h_1,h_2,h_3,k,k_i)$&$h_1+h_2-h_3+4k+6$&$(h_1,h_2,h_3)$&$(k,k_i)$&regular\\
(with $h_1 \geq h_2 \geq |h_3|$&&&&short\\
and $k \geq 0$)&&&&\\
&&&&\\
$c(h_1,h-\half,h+\half,k,k_i)$&$h_1+4k+\frac{11}{2}$&$(h_1-\half,h,h)$&$(k+\half,k_i)$&isolated\\ 
(with $h_1 \geq h+\half$&&&&short\\
$h \geq 0$ and $k \geq -\half$)&&&&\\
&&&&\\
$c(h,h,h+1,k,k_i)$&$h+4k+6$&$(h,h,h)$&$(k+1,k_i)$&isolated\\
(with $h \geq 0$ &&&&short\\
and $k \geq-1$)&&&&\\
&&&&\\
$c(-\half,-\half,\half,k,k_i)$&$4k+6$&$(0,0,0)$&$(k+\frac{3}{2},k_i)$&isolated\\
(with $k \geq -\frac{3}{2}$&&&&short\\
&&&&\\
        \hline
\end{tabular}
}

\subsection{Indices}\label{sec:indx}
As in the $d=3$ case, we define an index for $d=6$ as any linear combination 
of the multiplicities of short representations that evaluates to zero on every 
collection of representations that appear on the RHS of \eqref{cenl}, 
\eqref{cenm}, \eqref{cenn},and \eqref{ceno} so that it is invariant under 
any deformation of superconformal field theory under which the spectrum 
evolves continuously. We now proceed to list all of these indices,

\begin{enumerate}

\item The simplest indices are given by the multiplicities of 
short representations in the 
spectrum that never appear on the RHS of 
\eqref{cenl}, \eqref{cenm}, \eqref{cenn},and \eqref{ceno} 
(for any values of the quantum numbers on the LHS of those equations). 
 All such representations are easy to list; they are
\begin{itemize}
\item  
$c(h_1,h-\half,h+\half,k,k_i)$ for all $h_1 \geq h+\half$, $h \geq 0$  and $k-k_1=-\half,0$. 
\item
$c(h,h,h+1,k,k_i)$ for all $h \geq 0$ and $k-k_1=-1,-\half,0$
\item
$c(-\half,-\half,\half,k,k_i)$ for $k-k_1=-\frac{3}{2},-1,-\half,0$ 
\end{itemize}

In all the above cases we must consider all the possible values of the set 
${k_i}, i = 1 \ldots n-1$. This means $k_1 \geq k_2 \geq \ldots \geq k_{n-1} \geq 0$ and the $k_i$ may each be integers or half integers. 

\item We can also construct indices from linear combinations of the 
multiplicities of representations that do appear on the RHS 
of \eqref{cenl}, \eqref{cenm}, \eqref{cenn},and \eqref{ceno}. 
The complete list of such linear combinations is given by,

\begin{equation}\label{cveca} 
I_{M_1,M_2,M_3,\{k_i\}} =
\sum_{p=M_3-1}^{2(M_1-k_1)}(-1)^{p+1}n\{c(M_2+{p \over 2},{p \over
2},M_3-{p \over 2},M_1-{p \over 2},{k_i})\},
\end{equation}
where $n\{R\}$ denotes the number of representations of type $R$ and the Index 
labels $M_1$, $M_2$ and $M_3$ are respectively the values of  $h_2+k $, 
$h_1-h_2 $ and $M_3=h_2+h_3$ for the regular representations that 
appears in the above sum. Here $M_2$ and $M_3$ are integers greater than or equal 
to zero and $M_1$ is an integer or half integer with  $M_1 \geq {M_3 \over 2}+ k_1$.

\end{enumerate}

\subsection{Minimally BPS states: distinguished supercharge and commuting superalgebra}
Consider the special Q with charges $(h_1=-\half,h_2=-\half,h_3=\half,k=\half,\ep_0=\half)$. Let $S=Q^{\dagger}$; it is then easily verified that,

\begin{equation}\label{cdela} 
2\{S,Q\} \equiv \D = \ep_0 -(h_1+h_2-h_3+4k)
\end{equation}

Just as in $d=3$, we shall define a partition function over 
states annihilated by Q. Again all 
such states transform in an irreducible representation of the 
subalgebra of the superconformal algebra that commutes with $Q,S$ and hence 
$\Delta$. This subalgebra is easily determined to be the supergroup 
$D(3,\frac{n-2}{2})$ (see \cite{Minwalla:1997ka}).

The bosonic subgroup of this commuting superalgebra is $SU(3,1) 
\otimes Sp(n-2)$. The usual Cartan charges of $SU(3,1)$ and the Cartan charges 
of $Sp(n-2)$ are given in terms of the Cartan elements of the full 
superconformal algebra by,
\begin{equation}\label{crepa} 
 E=3\ep_0+h_1+h_2-h_3; H_1=h_1-h_2; H_2=h_2+h_3; K_i=k_{i+1},
\end{equation} 

where $E$ is the $U(1)$ Cartan, $(H_1,H_2)$ are the $SU(3)$ Cartans (in the Dynkin basis) and $K_i$ are the $Sp(n-2)$ Cartans (in the orthogonal
basis). \footnote{Specifically the Cartans $H_1$ and $H_2$ are the following $3\times 3$ SU(3) matrices,
\[ H_1=\left( \begin{array}{ccc}
        0 & 0 & 0 \\
        0 & 1 & 0 \\
        0 & 0 & -1 \end{array} \right),~~~ H_2=\left( \begin{array}{ccc}
        1 & 0 & 0 \\
        0 & -1 & 0 \\
        0 & 0 & 0 \end{array} \right)\].
}

\subsection{A Trace formula for the general Index and its Character
Decomposition}

As in the case of $d=3$, we define the Witten index as,

\begin{equation}\label{cwia} I^{W} = Tr_R[(-1)^F \exp{(-\zeta \D+\m G)}],
\end{equation}

Where the trace is evaluated over any Hilbert space that hosts a 
representation of the $d=6$ superconformal algebra. Here $F$ is the fermion 
number operator; by the spin statistics theorem, in any quantum field theory 
we take $F=2h_2$. G is any element of the subalgebra that commutes with 
the set set $\{ Q, S, \Delta \}$; by a similarity transformation , G may always be rotated in to a linear combination of the subalgebra Cartan generators.

The Witten index \eqref{cwia} receives contributions only from the states 
that are annihilated by both $Q$ and $S$ (all other states yields contribution 
that cancel in pairs) and, hence, have $\Delta =0$. So it is independent
of $\zeta$. The usual arguments\cite{Witten:1982df} also ensure that $I^W$ is also an index and 
hence it should be possible to expand $I^W$ as a linear combination of the 
indices defined in the previous section. In fact is easy to check that for any 
representation $A$ (reducible or irreducible) of the $d=6$ 
superconformal algebra,
    
\begin{equation}\label{cwif}
\begin{split}
&I^{wi}(A)=\sum_{M_1,M_2,M_3,\{k_i\} }I_{M_1,M_2,M_3}\ch_{sub}(M_2,M_3,k_i,4(M_2-M_3)+12M_1+24)\\
&+\sum_{\{k_i\},k-k_1=-\frac{3}{2},-1,-\half,0} n\{c(-\half,-\half,\half,k,k_i)\}\ch_{sub}(0,0,k_i,12k+18)\\
&+\sum_{\{k_i\},h\geq0,k-k_1=-1,-\half,0} (-1)^{2 h + 1} n\{c(h,h,h+1,k,k_i)\}\ch_{sub}(0,2h+1,k_i,4h+12k+20)\\
&+\sum_{\{k_i\},h_1,h(h_1\geq h\geq0),k-k_1=-\half,0}
(-1)^{2h}n\{c(h_1,h,h+1,k,k_i)\}\ch_{sub}(h_1-h,2h+1,k_i,4h_1+12k+20).
\end{split}
\end{equation}

where $\chi_{sub}(H_1,H_2,K_i,E)$ is the supercharacter of the representation with highest weights $H_1,H_2,K_i,E$ as defined in \eqref{crepa}. In the first sum 
in \eqref{cwif}
$M_2$ and $M_3$ run over integers greater than or equal 
to zero and $M_1$ runs over  integers or half integers with  $M_1 \geq {M_3 \over 2} + k_1$.
Also the set $\{k_i\}$ runs over integer and half integer values satisfying the condition $k_1 \geq k_2 \dots \geq k_n$.
In order to obtain \eqref{cwif} we have used,

\begin{equation}\label{cwib} 
\begin{split}
I^{wi}[(c(h_1,h_2,h_3,& k,k_i)(\rm{with}\  h_1 \geq h_2 \geq |h_3|\  \rm{and}\  k \geq 0)]=\\
&(-1)^{2h_2+1}\ch_{sub}(h_1-h_2,h_2+h_3,k_i,4(h_1+h_2-h_3)+12k+24).
\end{split}
\end{equation}
\begin{equation}\label{cwic}
\begin{split}
I^{wi}[(c(h_1,h,h+1,k,k_i)(\rm{with}&\  h_1 \geq h \geq 0 \ \rm{and}\  k \geq -\half)]=\\
&(-1)^{2h+1}\ch_{sub}(h_1-h,2h+1,k_i,4h_1+12k+20).
\end{split}
\end{equation}
\begin{equation}\label{cwid}
\begin{split}
I^{wi}[(c(h,h,h+1,k,k_i)(\rm{with}\  h \geq& 0 \ \rm{and}\  k \geq -1)]=\\
&(-1)^{2h+1}\ch_{sub}(0,2h+1,k_i,4h+12k+20).
\end{split}
\end{equation}
\begin{equation}\label{cwie}
\begin{split}
I^{wi}[(c(-\half,-\half,\half,k,k_i)(\rm{with}\  k \geq -\frac{3}{2})]=
\ch_{sub}(0,0,k_i,12k+18).
\end{split}
\end{equation}
Equations \eqref{cwib}-\eqref{cwie} follow from the observation that
the set of $\Delta = 0$ states (the only states
 that contribute to the Witten index) in any short representation of the 
superconformal algebra transform in a single representation of the commuting 
super subalgebra. The quantum numbers of these representations of the 
subalgebra are easily determined, given the quantum numbers of the parent
short representation.
 In the case of regular short representations, a primary of the 
subalgebra representation (in which the $\Delta =0$ states transform)is 
obtained by the acting on the  highest weight primary of the full 
representation (which turns out to have $\Delta=6$) with supercharges 
$Q_1,Q_2$ and $Q_3$ with quantum numbers $(h_1=\half,h_2=\half,h_3=\half,k=
\half,k_i=0,\ep_0=\half)$, $(h_1=\half,h_2=-\half,h_3=-\half,k=\half,k_i=0,\ep_0=
\half)$ and $(h_1=-\half,h_2=\half,h_3=-\half,k=\half,k_i=0,\ep_0=\half)$ 
respectively, all of which has $\Delta=-2$. The Witten index evaluated over 
these representations in terms of the supercharacter of the subgroup is given
 by \eqref{cwib}.

 In the case of isolated representations the highest weight primary of the  
full representation turns out to have $\Delta = 4,2$ and $0$; for the 
$\Delta = 4$ case  the primary of the subalgebra is obtained by the 
action of $Q_1$ and $Q_2$ on the primary of the full superconformal algebra,
 and for $\Delta = 2$ case it is obtained by the action of $Q_1$. The highest
 weight of an isolated superconformal short which itself has $\Delta = 0$
is also a primary of the commuting subalgebra. The Witten index 
evaluated over these representations in terms of the supercharacter of the 
subgroup is given by \eqref{cwic}, \eqref{cwid} and \eqref{cwie}.

Note that every index that appears in the list of subsection 
\S\S \ref{sec:indx} 
appears as the coefficient of a distinct subalgebra supercharacter 
in  \eqref{cwif}. As supercharacters of distinct irreducible representations 
are linearly independent,  it follows that knowledge of $I^{W}$ is sufficient 
to reconstruct all superconformal indices of the algebra. In this sense  
\eqref{cwif} is the most general index that can be  constructed  from 
the superconformal algebra alone.

\subsection{The Index over M theory multi gravitons in
$AdS_7 \times S^4$}

We now compute the Witten Index defined for the for the world volume theory
of the $M5$ brane in the large $N$ limit. The R-symmetry for this algebra is $SO(5)$ corresponding to rotations in the 5 directions transverse to the brane. This is consistent with the formalism above because $SO(5) \sim Sp(4)$. We will use the symbols $l_1, l_2$ to represent the $SO(5)$ Cartans in the orthogonal basis. The $Sp(4)$ Cartans are given by:
\begin{equation}
k = {l_1 + l_2 \over 2}, ~~~ k_1 = {l_1 - l_2 \over 2} 
\end{equation}
Note, also that the bosonic part of the commuting subalgebra is $SU(3,1) \otimes Sp(2)$. In the calculation below, we will us the equivalence $Sp(2) \sim SU(2)$. The $SU(2)$ charge is the same as the $Sp(2)$ charge.

In the strict large $N$ limit, the spectrum of this theory is the Fock space of supergravitons of $M$ theory on $AdS_7 \times S^4$ \cite{Maldacena:1997re,Aharony:1999ti}.\footnote{The index we will calculate is sensitive to ${1 \over 16}$ BPS states. However, the ${1 \over 4}$ BPS partition function has been calculated, even at finite $N$, in \cite{Bhattacharyya:2007sa}} The set of primaries for the graviton spectrum is 
($\ep_0=2p,l_1=2p,l_2=0,h_1=0,h_2=0,h_3=0$) \cite{Gunaydin:1984wc}\footnote{we specify the highest 
weight of the maximal compact subgroup; $\ep_0$ being the $SO(2)$ charge, 
$l_1$ and $l_2$ being the $SO(5)$ charges in orthogonal basis and $h_1,h_2$ 
and $h_3$ being the $SO(6)$ charge also in the orthogonal basis}, 
where p can be any positive integer.
Now given a highest weight state, we again use the procedure described in Appendix \ref{racahspeiser} 
to obtain the representations (of the maximal compact subgroup) occurring in 
the supermultiplet.
The result is enumerated in table~\ref{tb:d6gravspec} and agrees with \cite{Gunaydin:1984wc}. By the action of 
momentum operators on this states we can build up the entire 
representation of the superconformal algebra.

\TABLE[h]{
\caption{d=6 graviton spectrum}
\label{tb:d6gravspec}
\begin{tabular}{|c|c|c|c|c|c|}
	\hline         
range of $p$ & $\epsilon_0[SO(2)]$ & $SO(6)$[orth.] & $SO(5)$[orth.] & $\Delta$ & contribution  \\
	\hline
&&&&&\\
        
$p\geq1$   &$2p$            &$(0,0,0)$          &$(p,0)$             & $0$ & $+$ \\
$p\geq1$   &$2p+\frac{1}{2}$&$(\half,\half,\half)$      &$(\frac{2p-1}{2},\frac{1}{2})$            & $0$ & $+$ \\
$p\geq1$   &$2p+1$          &$(1,1,1)$          &$(p-1,0)$        & $2$ & $+$ \\
$p\geq2$   &$2p+1$          &$(1,0,0)$&$(p-1,1)$    & $0$ & $+$ \\
$p\geq2$   &$2p+\frac{3}{2}$&$(\frac{3}{2},\half,\half)$  &$(\frac{(2p-3)}{2},\frac{1}{2})$     &$2$ & $+$ \\
$p\geq2$   &$2p+2$          &$(2,0,0)$         &$(p-2,0)$            & $4$ & $+$ \\
$p\geq3$   &$2p+\frac{3}{2}$&$(\half,\half,-\half)$      &$(\frac{2p-3}{2},\frac{3}{2})$        & $0$ & $+$ \\
$p\geq3$   &$2p+2$          &$(1,1,0)$          &$(p-2,1)$& $2$ & $+$ \\
$p\geq3$   &$2p+\frac{5}{2}$&$(\frac{3}{2},\half,-\half)$&$(\frac{(2p-5)}{2},\frac{1}{2})$ & $4$ & $+$ \\
$p\geq3$   &$2p+3$          &$(1,1,-1)$      &$(p-3,0)$    & $6$ & $+$ \\
$p\geq4$   &$2p+2$          &$(0,0,0)$      &$(p-2,2)$& $2$ & $+$ \\  
$p\geq4$   &$2p+\frac{5}{2}$&$(\half,\half,\half)$          &$(\frac{2p-5}{2},\frac{3}{2})$& $4$ & $+$ \\
$p\geq4$   &$2p+3$          &$(1,0,0)$          &$(p-3,1)$        & $6$ & $+$ \\
$n\geq4$   &$2p+\frac{7}{2}$&$(\half,\half,-\half)$          &$(\frac{2p-7}{2},\half)$    & $8$ & $+$ \\  
$p\geq4$   &$2p+4$          &$(0,0,0)$          &$(p-4,0)$& $12$ & $+$ \\
        
	\hline
        
$p=1$      &   $\frac{7}{2}$         &$(\half,\half,-\half)$      &($\half,\half$) & $0$ & $-$ \\
$p=1$      & $4$ &$(1,1,0)$          &$(0,0)$& $2$ & $-$ \\
$p=1$ &$4$ &$(0,0,0)$ & $(1,0)$ & $2$ & $-$ \\
$p=1$ &$5$  &$(1,0,0)$ &$(0,0)$ & $4$ & $+$\footnotemark

 \\
$p=1$ &$6$  &$(0,0,0)$ &$(0,0)$ & $6$ & $-$\\
        \hline 
$p=2$      &   $6$         &$(0,0,0)$          &$(1,1)$    & $2$ & $-$ \\
$p=2$      & $\frac{13}{2}$ &$(\half,\half,\half)$  &$(\half,\half)$                 & $4$ & $-$ \\
$p=2$      & $7$           &$(1,0,0)$          &$(0,0)$                 & $6$ & $-$ \\
        \hline
        \hline

\end{tabular}

}
\footnotetext{The `+' appears 
because the conformal representation we subtract is, itself short.
 See \cite{Barabanschikov:2005ri} for details}

It is now again simple to compute the Index over single gravitons once we have 
the spectrum. The Witten Index for the $p^{th}$ graviton representation
($R_p$)(i.e. for a particular value of p in the primary), is obtained by

\begin{equation}\label{chip2} \begin{split}I^W_{R_p} =& Tr_{\Delta = 0} \Big[ 
{(-1)^F x^E z^{K_1} y_1^{H_1} y_2^{H_2}}
\big]\\
=& \sum_{q}\frac{(-1)^{2(h_2)_q} x^{(3\epsilon_0+h_1+h_2-h_3)_q} \chi^{SU(2)}_q(z)\chi^{SU(3)}_q(y_1,y_2)}
{(1-x^4y_1)(1-\frac{x^4y_2}{y_1})(1-\frac{x^4}{y_2})}
,\end{split}
\end{equation}

where $q$ runs over all the conformal representations with $\Delta=0$ that 
appears in the decomposition of $R_p$ in table~\ref{tb:d6gravspec}; 
$x,z,y_1$ and $y_2$ are the exponential of the chemical potentials 
corresponding to the subgroup charges $E,K_1,H_1$ and $H_2$ respectively as 
defined in \eqref{crepa}; $\chi^{SU(2)}$ and $\chi^{SU(3)}$ denote the 
characters of the groups $SU(2)$ and $SU(3)$ respectively, which are computed 
using the Weyl character formula.

The Index over the single particle states  is then simply given by 
the following sum,
\begin{equation}\label{indspa2} \begin{split}
I^W_{sp} = \sum_{p=3}^{\infty}I^W_{R_p}+I^W_{R_2}+I^W_{R_1},
\end{split}
\end{equation} 
Performed this sum, we find that the single particle contribution to the
index is
\begin{equation}\label{indspb6} 
\begin{split}
I^W_{sp} &= {{\rm term1} + {\rm term2} \over {\rm den}} \\
{\rm term1} &= x^6 \left(\sqrt{z} y_1^2 \left(1-x^8 y_2\right)
   x^2+\sqrt{z} y_2 \left(1-x^8 y_2\right) x^2 \right) \\
{\rm term2} &= x^6 \left(y_1 \left(-\sqrt{z} x^{10}+\sqrt{z} y_2^2
   x^2+\left(x^{12}-1\right) (z+1)
   y_2\right)\right) \\
{\rm den} &= \left(\sqrt{z} x^{12}-(z+1)
   x^6+\sqrt{z}\right) \left(x^4 y_1-1\right)
   \left(x^4-y_2\right) \left(x^4 y_2-y_1\right)
.\end{split}
\end{equation}

The index over the Fock-space of gravitons can be obtained from the above 
single particle index by the formula \eqref{indfs}.

To get a sense for the formula, let us set $z, y_i \rightarrow 1$ in \eqref{indspb6} leaving only  $x \equiv e^{-\beta}$. We remind the reader that
$\beta$ is the chemical chemical potential corresponding to $E = 3\epsilon_0+h_1+h_2-h_3$. 
This leads to
\begin{equation}\label{indspx} \begin{split}\left. I^W_{sp}(x) \right|_{z, y_i \rightarrow 1}=
\frac{x^6 \left(2
   x^4+x^2+2\right)}{\left(x^8+x^6-x^2-1\right)^2}
.\end{split}
\end{equation}

We note that in the high energy limit when $x \ra 1$,  
$I^W_{sp}$ in \eqref{indspx} becomes $I^W_{sp} = \frac{5}{144\beta^2}$. 
Then by the use of \eqref{indfs} we have,

\begin{equation}
I^W_{fock}=\exp{\frac{5\zeta(3)}{144\beta^2}}.
\end{equation}

Then the average value of $E = 3\epsilon_0+h_1+h_2-h_3$ is given by,
\begin{equation}  E=-{\frac{\partial \ln I^W_{fock}}{\partial \beta}}  =  {5\zeta(3) \over 72\beta^3}.
\end{equation}

If we define an entropy like quantity $S$ by
\begin{equation}
I^W_{fock} = \int dy \exp{(-\beta y)} \exp{S^{\rm ind}(y)},
\end{equation} 
we find,
\begin{equation} 
S^{\rm ind}(E) =\frac{5\zeta(3)/48}{(5\zeta(3)/72)^{\frac{2}{3}}} E^{2 \over 3}.
\end{equation}

We can also do a similar analysis with the partition function instead of the 
index. The single particle partition function evaluated on the $\Delta = 0$ 
states with all the other chemical potentials except $\beta$ set to zero is 
given by,
\begin{equation}\label{ptfnsp2} \begin{split}
Z_{sp}(x) ={\rm tr}_{\Delta=0} x^{E} = \frac{-x^6 \left(-2
   x^8+x^6+x^2-2\right)}{\left(1-x^2\right)^5
   \left(x^2+1\right) \left(x^4+x^2+1\right)^2}
\end{split}
\end{equation}
The separate contribution of the bosonic and fermionic states to the partition function 
in \eqref{ptfnsp2} are as follows,
\begin{equation}
Z_{\rm sp}^{\rm bose}(x)={\rm tr}_{\Delta=0~{\rm bosons}}= \frac{x^6 \left(3 x^{10}-x^6+2\right)}{\left(1-x^4\right)^3
   \left(1-x^6\right)^2}
\end{equation}
\begin{equation}
Z_{\rm sp}^{\rm fermi}(x) ={\rm tr}_{\Delta=0~{\rm fermions}} = \frac{x^8 \left(2 x^{10}-x^4+3\right)}{\left(1-x^4\right)^3
   \left(1-x^6\right)^2}
\end{equation}

An analysis similar to that done for the Index, yields for the above
partition function 
\begin{equation}
\ln Z_{\rm fock} = \sum_n {Z_{\rm sp}^{\rm bose}(x^n) + (-1)^{n+1}Z_{\rm sp}^{\rm fermi} \over n} = \frac{7\zeta(6)}{2048 \beta^5}
\end{equation}

\begin{equation} 
S(E) = \frac{21\zeta(6)/1024}{(35\zeta(6)/2048)^{\frac{5}{6}}} E^{5 \over 6},
\end{equation}
which is again similar to that of a six dimensional gas for reasons that are
similar to those explained below equation \eqref{m2susygravgrowth}. Note, that in this case, we have 2 transverse supersymmetric scalars and 3 derivatives.

\subsection{The Index on the worldvolume theory of a single $M5$ brane}
We will now compute our index over the worldvolume theory of a single $M5$ brane. For this free theory, the single particle state content is just the representation corresponding to $p=1$ in Table \ref{tb:d6gravspec} of the previous subsection. This means that it corresponds to the representation of the $d=6$ superconformal group with the primary having charges $\epsilon_0 = 2 $, $SO(6)$ highest weights $[0,0,0]$ and R-symmetry $SO(5)$ highest weight $[1,0]$. 
Physically, this multiplet corresponds to the 5 transverse scalars, real fermions transforming as chiral spinors of both $SO(6)$ and $SO(5)$ and a self-dual two form $B_{\mu \nu}$. See \cite{Claus:1997cq,Seiberg:1997ax,Minwalla:1998rp} and references therein for more details. 
Using Table \ref{tb:d6gravspec}, we calculate the Index over these states
\begin{equation}
\label{m5index}
\begin{split}
I_{M_5}^{\rm sp}(x,z,y_1,y_2) &=  {\rm Tr} \left[(-1)^F x^E z^{ K_1} y_1 ^{H_1} y_2^{H_2} \right]  \\
&= {x^{6}(z^{1 \over 2} + {1 \over z^{1 \over 2}})  -  x^8 \left(y_2 + {y_1 \over y_2} + {1 \over y_1}\right) + x^{12}  \over \left(1 - x^4 y_1 \right) \left(1 - x^4 {y_2  \over y_1} \right) \left(1 - {x^4 \over y_2} \right)} 
\end{split}
\end{equation}
Specializing to the chemical potentials $y_i \rightarrow 1, z \rightarrow 1$, the index simplifies to
\begin{equation}
I_{M_5}^{\rm sp}(x,z=1,y_i=1) = {2 x^6 - 3 x^8 + x^{12} \over (1 - x^4)^3}
\end{equation}
Multiparticling this index, to get the index over the Fock space on the $M_2$ brane, we find that 
\begin{equation}
\begin{split}
I_{M_5}(x,z=1, y_i = 1) &= \exp{\sum_n {I_{M_5}^{\rm sp}(x^n,z=1,y_i = 1) \over n}}\\
&= \prod_{n_1,n_2,n_3} {\left(1 - x^{8 + 4 (n_1+n_2+n_3)}\right)^3 \over \left(1 - x^{6 + 4(n_1 + n_2 + n_3)} \right)^2 \left(1 - x^{12 + 4 (n_1 + n_2 + n_3)} \right)}
\end{split}
\end{equation}
At high temperatures $x \equiv e^{-\beta} \rightarrow 1$, we find
\begin{equation}
\left.I_{M_5}\right|_{x \rightarrow 1, y_i = 1} = \exp\left\{{\pi^2 \over 32 \beta}\right\}
\end{equation}

The supersymmetric single particle partition function, on the other hand is given by
\begin{equation}
\begin{split}
Z_{M_5}^{\rm sp, susy}(x,z,y_1,y_2) &=  {\rm Tr}_{\Delta = 0} \left[x^E z^{ K_1} y_1 ^{H_1} y_2^{H_2} \right] \\
&=  {x^{6}(z^{1 \over 2} + {1 \over z^{1 \over 2}})  +  x^8 \left(y_2 + {y_1 \over y_2} + {1 \over y_1}\right) + x^{12}  \over \left(1 - x^4 y_1 \right) \left(1 - x^4 {y_2  \over y_1} \right) \left(1 - {x^4 \over y_2} \right)} 
\end{split}
\end{equation}
In particular, setting $z,y_i = 1$, we find
\begin{equation}
Z_{M_5}^{\rm sp, susy}(x,z=1,y_i=1) = {2 x^6 + 3 x^8 + x^{12} \over (1 - x^4)^3}
\end{equation}
with contributions from the bosons and fermions being
\begin{equation}
\begin{split}
Z_{M_5}^{\rm sp, susy, bose}(x) &= {\rm tr}_{\Delta=0~{\rm bosons}} x^{E} = {2 x^6 + x^{12} \over (1 - x^4)^3} \\
Z_{M_5}^{\rm sp, susy, fermi}(x) &= {\rm tr}_{\Delta=0~{\rm fermions}} x^{E} = {3 x^8  \over (1 - x^4)^3} 
\end{split}
\end{equation}
Multiparticling this result, we find 
\begin{equation}
\begin{split}
&Z_{M_5}(x,z=1, y_i = 1) = \exp{\sum_n {Z_{M_5}^{\rm sp, susy}(x^n,z=1,y_i = 1) \over n}}\\
&= \prod_{n_1,n_2,n_3} { \left(1 + x^{8 + 4 (n_1+n_2+n_3)} \right)^3 \over \left(1 - x^{6 + 4(n_1 + n_2 + n_3)} \right)^2 \left(1 - x^{12 + 4 (n_1 + n_2 + n_3)} \right)}
\end{split}
\end{equation}
At high temperatures $x \rightarrow 1$, we find that
\begin{equation}
Z_{M_5}(x \rightarrow 1, z=1, y_i = 1) \approx \exp\left\{{45 \zeta(4) \over 512 \beta^3}\right\}
\end{equation}

\section{d=5}
\label{d5section}
\subsection{The Superconformal Algebra and its Unitary Representations}
In $d=5$, the bosonic part of the superconformal algebra is $SO(5,2) \otimes SU(2)$.
Under the $SO(5) \otimes SO(2)$ subgroup of the conformal group the
supersymmetry generators $Q^i_{\m}\ i = 1,\cdots,4$ and $\m =
\pm\half$ transform as the spinors of $SO(5)$, with charge $\half$
under $SO(2)$. The R-symmetry group is $SU(2)$  and $\mu$ above is an $SU(2)$ index. We use $k$ to represent the $SU(2)$ Cartan.
The $SO(5)$ Cartans in the orthogonal basis are denoted by $h_1, h_2$. 
We will use $\epsilon_0$ to represent the energy which is measured by the charge under $SO(2)$. To lighten the notation, we will use the same symbols to represent the eigenvalues of states under these Cartans. 

With these conventions the $Qs$ have $\epsilon_0={1 \over 2}, k = \pm {1 \over 2}$ and $SO(5)$ charges:
 \begin{equation}\label{char}
\begin{split}{Q^1\rightarrow& (\half,\half), ~~~
                    Q^2\rightarrow (\half,-\half)\\
                    Q^3\rightarrow& (-\half,\half), ~~~
                    Q^4\rightarrow (-\half,-\half)}\end{split}
                    \end{equation}
The superconformal generators $S_i^{\m}$ are the conjugates of $Q^i_{\m}$ and therefore their charges are the negative of the charges above.

The anticommutator between $Q$ and $S$ is given by \begin{equation}\label{five}
\{S_i^{\m},Q^j_{\n}\} \sim \delta_{\n}^{\m}\left(T_i^j\right) - \delta_i^jM_\n^\m
\end{equation} 
Here  $T_i^j$ and $M_\n^\m$ are the $SO(5,2)$ and $SU(2)$
generators respectively.

As in the previous sections, by diagonalizing this operator one can determine when a descendant of the primary will have zero norm. Performing this analysis \cite{Minwalla:1997ka}, one finds that short representations can exist when the highest weights of the primary satisfy one of the following conditions
\begin{equation}
\label{ssh2} 
\begin{split}
\epsilon_0 =& h_1 + h_2 + 3k + 4 \ \ \rm{when}\ h_1 \geq h_2 \geq 0 \ \rm{and}~ k \geq 0,\\
\epsilon_0 =& h_1 + 3k + 3,\ \ \rm{when}\ h_2 = 0\ \rm{and}~ k \geq 0,\\
\epsilon_0 =& 3k,\ \ \ \rm{when}\  h_1 = h_2 = 0,\ \rm{and}~ k \geq 0.
\end{split}
\end{equation}
The last two conditions give isolated short representations.

\subsection{Null Vectors and Character Decomposition of a Long Representation at the Unitarity Threshold}
As in the case of $d=3,6$, and as explained in the previous section the 
short representations of $d=5$ are also either {\it regular} or 
{\it isolated}. The energy of a {\it regular} short representation is 
given by $\epsilon_0 = h_1+h_2+3k+4$. Again the null states of such a 
representation transform in an irreducible representation of the algebra; for $h_1 \neq 0 \neq h_2$
the highest weight of the primary at the head of this null irreducible representation is given 
in terms of the highest weight of the primary of the representation itself 
by $\epsilon_0'=\epsilon_0+\half,~~~k'=k+\half,~~~h_1'=h_1-\half,
~~~h_2'=h_2-\half$. We note that $\epsilon_0'-h_1'-h_2'-3k'-4=
\epsilon_0-h_1-h_2-3k-4=0$, which shows that the null states also transform 
in a {\it regular} short representation. Thus a long representation at the 
edge of this unitarity bound has the same state content as the union of 
ordinary and null states of such a {\it regular} short representation.
So we conclude that,
\begin{equation}\label{beng}
\begin{split}
\lim_{\delta \ra 0}\ch(h_1+h_2+3k+4+\delta,h_1,h_2,k] =&\ch(h_1+h_2+3k+4,h_1,h_2,k)\\
                                        &+ \ch(h_1+h_2+3k+\frac{9}{2},h_1-\half,h_2-\half,k+\half),\\
                                        &(\rm{with}\ h_1 \geq h_2 \geq {1 \over 2} \ \rm{and}\  k \geq 0).
\end{split}
\end{equation}
where $\chi(\epsilon_0,h_1,h_2,k)$ is the character of the irreducible representation with energy
$\epsilon_0$, $SO(5)$ highest weights(in the orthogonal basis) $(h_1,h_2)$ and $SU(2)$ highest weight
$k$.

Now when $h_1 \geq 1,h_2=0$ the null states of the regular short 
representation occur at level two and are characterized by a primary 
with the highest weights $\epsilon_0'=\epsilon _0+1,~~~k'=k+1,~~~h_1'=h_1-1,~~~
h_2'=0$. Now we note that $h_1'\neq0,h_2'=0$ and $\epsilon_0'-h_1'-3k'-3=
\epsilon_0-h_1-3k-4=0$, and so we conclude that the null states of such a 
type of {\it regular} short representation transform in an {\it isolated}
short representation. Thus for a long representation at the edge of such 
a unitarity bound we have,
\begin{equation}\label{bengm}
\begin{split}
\lim_{\delta \ra 0}\ch(h_1+3k+3+\delta,h_1,h_2=0,k) =& \ch(h_1+3k+3,h_1,h_2=0,k)\\
                                                           &+ \ch(h_1+3k+4,h_1-1,h_2=0,k+1).\\ &h_1 \geq 1, k \geq 0
\end{split}
\end{equation}

Finally when $h_1=0=h_2$ the null states of the {\it regular} short 
representation occur at level four and are labeled by a primary with the 
highest weight $\epsilon_0'=\epsilon_0+2,~~~k'=k+2,~~~h_1'=0,~~~h_2'=0$.
Here we note that $h_1'=0=h_2'$ and $\epsilon_0'-3k'=\epsilon_0-3k-4=0$, 
which shows that the null states of this type of {\it regular} short 
representation again transforms in an {\it isolated} short representation but 
the {\it isolated} short representation encountered here is different from the
one encountered in the previous paragraph. Thus for long representations at 
the edge of this unitarity bound we have,

\begin{equation}\label{benh} 
\lim_{\delta \ra 0}\ch(3k+\delta,h_1=0,h_2=0,k) = \ch(3k,0,0,k) +
\ch(3k+2,0,0,k+2), ~~ k \geq 0.
\end{equation}

Thus we see that the  {\it isolated} short representations (as defined in the previous subsection) 
are separated from other representations with the same $SO(5)$ and $SU(2)$ weights by a finite gap in 
energy so it is not possible to {\it approach} such representations with long 
representations and therefore we do not have any equivalent of \eqref{bengm} 
or \eqref{benh} at energies near $h_1=3k+3$ (when $h_1\geq1,h_2=0$) 
or near $3k$ (when $h_1=0=h_2$) with $k \geq 0$ in both the cases.

For use below we define the following notation. Let $c(h_1,h_2,k)$ denote a 
regular short representation with $SO(5)$ and $SU(2)$ highest weights 
$(h_1,h_2)$ and $k$ respectively, and with $\epsilon_0 = h_1+h_2+3k+4$ (when
$h_1 \geq h_2 \geq 0$). We now extend this notation to include isolated short representations.
\begin{itemize}
\item $c(h_1,-\half,k)$ with $h_1>0$ and $k \geq -\half$ denotes the 
representation with SO(5) weights $(h_1-\half,0)$ and SU(2) 
quantum number $k+\half$ and with $\epsilon_0=h_1+3k+4$.
\item $c(-\half,-\half,k)$ with $k \geq -\frac{3}{2}$ denotes the representation with SO(5) weights $(0,0)$ and SU(2) 
quantum number $k+\frac{3}{2}$ and $\epsilon_0=3k+\frac{9}{2}$.
\end{itemize}

\subsection{Indices}
\label{sec:indx5}
As in the previous cases of $d=3,6$ for $d=5$ an Index is defined to be any linear combination 
of multiplicities of short representations that evaluates to zero on every 
collection of collection of representations that appears on the RHS of 
\eqref{beng}, \eqref{bengm} and \eqref{benh}.We now list these 
Indices.

\begin{enumerate}
\item{The multiplicities of short representations which 
never appear on the R.H.S of \eqref{beng}, \eqref{bengm}
and \eqref{benh}. These are  $c(-\half,-\half,k)$ for 
$k=0,-\half,-1,-{3 \over 2}$ and $c(h_1,-\half,k)$ for all $h_1 > 0$ and
$k=0,-\half$.}

\item{The complete list of indices constructed from linear 
combinations of the multiplicities of representations that appear
on the RHS of \eqref{beng}, \eqref{bengm} and \eqref{benh} is given by,
\begin{equation}\label{bveca} I^{(1)}_{M_1,M_2} =
\sum_{p=-1}^{2M_2}(-1)^{p+1}n\{c(M_1+{p \over 2},{p \over
2},M_2-{p \over 2})\},
\end{equation}
where $n\{R\}$ denotes the multiplicities of representations of type $R$, and 
the Index label $M_1$ and $M_2$ are the values of $h_1-h_2=
h_1-{p\over 2}$ and $h_2+k={p \over 2}+k$ for every regular representation 
that appears in the sum above. Here $M_1$ can be a integer greater than or equal to zero
and $M_2$ is an integer or  half integer greater than or equal to zero.
}
\end{enumerate}

\subsection{Minimally BPS states: distinguished supercharge and commuting superalgebra}

We consider the special Q with charges $(h_1=-\half,h_2=-\half,k=\half,
\epsilon_0=\half)$. Let $S=Q^{\dagger}$ then we have,

\begin{equation}\label{bdela} 
\D \equiv \{S,Q\} = \epsilon_0 -(h_1+h_2+3 K)
\end{equation}
We are now interested in a partition function over states annihilated by
this special $Q$. Such states transform in an irreducible representation of the subalgebra of the 
superconformal algebra that commutes with $\{Q,S,\D\}$. This subalgebra turns 
out to be $SU(2,1)$. Note that unlike $d=3,6$  this subalgebra is a bosonic lie algebra, and not a super lie algebra.
In the subalgebra, we will label states by their weights under the Cartan elements $H_1^s, H_2^s$, which are defined in terms of the Cartans of the full algebra by:
\begin{equation}\label{brepa} H_1^s=h_1-h_2,~~~
H_2^s=\epsilon_0+{h_1+h_2 \over 2}.
\end{equation}
Here, $h_1, h_2$ are the Cartans of the $SO(5)$ algebra in the orthogonal basis and $\epsilon_0$ represents the charge under $SO(2)$. 

\subsection{A Trace formula for the general Index and its Character
Decomposition}

We define the Witten Index,
\begin{equation}\label{bwia}
 I^{w} = {\rm Tr}_R[(-1)^F \exp(-\zeta\D+\m G)],
\end{equation} 
where the trace being evaluated over any Hilbert Space that hosts a reducible
or irreducible representation of the $d=5$ superconformal algebra.
Here $G$ is any element of the subalgebra that commutes with 
the set $\{S,Q,\D\}$ and  $F=2h_1$.  
It is always possible to express $G$
as a linear combination of the subalgebra Cartans (as given by 
\eqref{brepa}) by a similarity transformation. Once again, the Witten 
Index is independent of $\zeta$.
 
It is easy to check that the Witten Index $I^{W}$ evaluated on any 
representation $A$ (reducible or irreducible) is given by,
\begin{equation}\label{bwie} \begin{split}I^{W}(A)=&
\sum_{M_1,M_2}I^{(1)}_{M_1,M_2}\ch_{sub}(M_1,{3 \over
2}M_1+3(M_2+2)) \\&+
\sum_{h_1(\geq \half);k=-\half,0}
n\{c(h_1,-\half,k)\}\ch_{sub}(h_1+\half,\frac{3}{2}h_1+3k+\frac{21}{4})
\\&+ \sum_{k=-{3 \over
2},-1,-\half,0}n\{c(-\half,-\half,k)\}\ch_{sub}(0,3k+\frac{9}{2})
\end{split}
\end{equation}

with $\chi_{sub}(H_1^s,H_2^s)$ is the character of a representation of the subgroup, with highest weights $(H_1^s,H_2^s)$ in the conventions described above.

In order to obtain \eqref{bwie} we have used,
\begin{equation}\label{bwib} I^{wi}(c(h_1,h_2,k))=
(-1)^{2h_2+1}\ch_{sub}(h_1-h_2,\frac{3}{2}(h_1+h_2)+3k+6)
\end{equation}
\begin{equation}\label{bwic} I^{wi}(c(h_1,-\half,k))=\ch_{sub}
(h_1+\half,\frac{3}{2}h_1+3k+\frac{21}{4})
\end{equation} \begin{equation}\label{bwid}
I^{wi}(c(-\half,-\half,k))=\ch_{sub}(0,3k+\frac{9}{2})
\end{equation}

Note that the states with $\D = 0$ in any short representation (which are 
the states that contribute to the Witten Index), may be 
organized into a single irreducible representation of the subalgebra that 
commutes with $Q$.  The quantum numbers of this subalgebra representation
may be determined in terms of the quantum numbers of the parent short 
representation.  For a {\it regular} short representation the 
primary of the full  representation has $\D=4$ so the
 highest weight state of the representation of the subalgebra is 
reached by acting on it with the supercharges  $Q_1,Q_2,Q_3$ with
the charges $(h_1=\half,h_2=\half,k=\half,\epsilon_0=\half)$,
$(h_1=\half,h_2=-\half,k=\half,\epsilon_0=\half)$,
$(h_1=-\half,h_2=\half,k=\half,\epsilon_0=\half)$. These have $\D=-2,-1,-1$ 
respectively. Similarly an {\it isolated} short representation of type
$c(h_1,-\half,k)$ with $h_1 > 0$ and $k \geq -\half$ has $\D=3$ and is acted upon by $Q_1$ and 
$Q_2$ in order to reach the 
highest weight state of the representation of the subalgebra. Finally the {\it isolated} short 
representations of type $c(-\half,-\half,k)$ with $k \geq -\frac{3}{2}$ have $\D=0$ and are themselves the highest weight states of the representation of the 
subalgebra.

We finally note that every Index constructed in subsection \S\S \ref{sec:indx5}
appears as the coefficient of a distinct subalgebra character in \eqref{bwie}.
Thus $I^{W}$ may be used to reconstruct all superconformal Indices of the 
algebra which makes it the most general Index that is possible to construct 
from the algebra alone. 

\section{Discussion}
In this paper we have presented formulae for the most general superconformal Index for superconformal algebras
in 3, 5 and 6 dimensions. Our work generalizes the analogous construction of an index for four dimensional
conformal field theories presented in \cite{Kinney:2005ej}.

We hope that our work will find eventual use in the study of the space of superconformal field theories in 3, 5 and 6 dimensions. 
It has recently become clear that the space of superconformal field theories in four dimensions is
much richer than previously suspected \cite{Intriligator:2004xi}. 
The space of superconformal field theories in $d=3,5, 6$ may be equally intricate, although this
question has been less studied.  As our index is constant on any
connected component in the
space of superconformal field theories, it may play a useful role in
the study of this space.

In this paper we have also demonstrated that the most general superconformal index, in all the dimensions that
we have studied, is captured by a simple trace formula. This observation may turn out to be useful as traces may
easily
be reformulated as path integrals, which in turn can sometimes be evaluated, using either perturbative techniques
or localization arguments.

The two dimensional index -- the elliptic genus -- has played an important role in the understanding of black hole
entropy from string theory. However the four dimensional index defined in \cite{Kinney:2005ej} does not seem to capture the entropy
of black holes in any obvious way. It would be interesting to know what the analogous situation in in 3 and 6
dimensions. It would certainly be interesting, for instance, if the index for the theory on the world volume of
the
M2 of M5 brane underwent a large $N$ transition as a function of chemical potentials, to a phase whose index
entropy scales like $N^{\frac{3}{2}}$ and $N^{3}$ respectively. As we currently lack a computable framework
for multiple M2 or M5 branes we do not know if this happens; however see \cite{Bagger:2007vi} for recent interesting progress in this
respect.

In this connection we also note that the index for the weakly coupled Chern Simons theories studied in this paper
does undergo a large $N$ phase transition as a function of temperature. It would be interesting to have a holographic dual description of these phase transitions.

\section*{Acknowledgements}
The work of SM was supported in part by a Swarnajayanti Fellowship. We would like to thank L. Grant, S. Lahiri, R. Loganayagam, S. Nampuri, S. Wadia and all the students in the TIFR theory room for useful discussions. 

\section*{Appendices}
\appendix

\section{The Racah Speiser Algorithm}
\label{racahspeiser}
In this appendix, we describe the Racah-Speiser algorithm, that may be used to determine the state content of the supergraviton representations described in Tables \ref{tb:d3gravspec} and \ref{tb:d6gravspec}. This appendix is out of the main line of this paper, since this state content may also be found in \cite{Gunaydin:1985tc,Gunaydin:1984wc,biran1984fss}

First, we remind the reader how irreducible representations of Lie Algebras, and affine Lie Algebras may be constructed using Verma modules \cite{DiFrancesco:1997nk,fuchs1997sla}. A nice description that is particularly applicable to our situation is provided in \cite{Dobrev:2004tk}.

One starts by decomposing the algebra (${\cal G}$) as :
\begin{equation}
{\cal G} =  {\cal G}^{+} \oplus {\cal H} \oplus {\cal G}^{-}
\end{equation}
where ${\cal G}^{+}$(${\cal G}^{-}$) corresponds to the positive (negative) roots of ${\cal G}$ and ${\cal H}$ is the Cartan subalgebra. 

To construct the Verma module ${\cal V}$ corresponding to a lowest weight $|\Omega \rangle$,  one considers the linear space made up of the states $P({\cal G}^{+})  |\Omega \rangle$ where $P$ is any polynomial of the positive generators.

One may calculate the character of this module,
\begin{equation}
\chi_{\cal V} ({\bf \mu}) = {\rm tr}_{\cal V} e^{{\bf \mu} \cdot {\cal H}}
\end{equation}
where ${\mu}$ is a vector in the dual space of ${\cal H}$. The Weyl group ${\cal W}$ of the algebra has a natural action on ${\cal H}$ and this induces a natural action on ${\mu}$. Finally, to obtain the character of the irreducibe representation  
 $R(\Omega)$, one symmetrizes $\chi_{\cal V}$ with respect to ${\cal W}$.
\begin{equation}
\chi_{R(\Omega)}=\sum_{w \in {\cal W}} \chi_{\cal V}(w({\bf \mu})).
\end{equation}
One may now read off the list of states in $R$ using $\chi_{R(\Omega)}$.

Let us elucidate the method above by constructing the character of a representation of $SU(2)$ of weight $j$. 
If $J_{\pm}$ denote the raising and lowering operators and $J_3$ be the Cartan, then the Verma module corresponding to a lowest weight state of weight $|-j \rangle$ is spanned by the states $(J_+)^l|-j\rangle$ with $l=0,1,2,\ldots $. The character for this Verma module is given by,
\begin{equation}
(\chi_{\cal V})_j(x)= {\rm tr}x^{J_3} = \sum_{l=0}^{\infty}x^{-j+l} = \frac{x^{-j+1}}{1-x},
\end{equation}
The Weyl group of $SU(2)$ is $\mathbb{Z}_2$ which has two elements. One is just the identity. The other takes $x \rightarrow x^{-1}$. So the character of the irreducible representation corresponding to the highest weight $j$ is given by
\begin{equation}
\chi_j(x)={x^{j+1} \over x -1} + {x^{-j - 1} \over x^{-1} - 1} = \frac{x^{j+\half}-x^{-j-\half}}{x^{\half}-x^{-\half}}.
\end{equation}

The corresponding theory for superalgebras is not as well known but was developed following the work of Kac \cite{kac1978rcl}. Its application to the 4 dimensional superconformal algebra may be found in \cite{Dobrev:2004tk,Dolan:2002zh}. Here, although we do not have a proof of this algorithm from first principles, we have followed the natural generalization of the procedure described in \cite{Dolan:2002zh,Dolan:2005wy,Bianchi:2006ti} for superconformal algebras in $d=4$.

Starting with a lowest weight state one acts on this state with all the `raising' operators of the algebra(which includes the supersymmetry generators). Then, one discards null states and all their descendants as explained in the sections above. This process results in a Verma module. 

The character of this Verma module is particularly easy to construct. Although the exact structure of null vectors may be quite complicated as, for example, in section \ref{nullstructure6d}, the charges characterizing the null state (which is all that is important for the character) are always obtained by adding the charges of a particular supercharge (or combination of supercharges) to the charges of the primary. So, the character of the Verma module may be obtained by counting all possible actions of supercharges except for the specific combinations that lead to null states or their descendants. 

One now symmetrizes this character over the Weyl group of the {\em maximal compact subgroup} to obtain the character of the irreducible representation corresponding to our highest weight. 

\section{Charges}
\label{appcharges}
In this appendix, explicitly list the charges of the 
supersymmetry generators in the worldvolume theory of the $M2$ and $M5$ branes 
and also for the superconformal algebra in $d=5$. 
For the $M2$ brane, we have $16$ supersymmetry generators `Q'. We use the notation [$\epsilon_0,j,h_1,h_2,h_3,h_4$], where  $\epsilon_0$ is the energy, $j$ the $SO(3)$ charge and $h_1,h_2,h_3,h_4$ are the $SO(8)$ charges in the orthogonal basis (with a choice of Cartans in which the Qs are in the vector). With this notation, the Qs have charges
\begin{equation}\label{dIIIQs}\begin{split}Q_1 &=[\half,\half,1,0,0,0]\   ;\   Q_2=[\half,\half,-1,0,0,0],\\Q_3&=[\half,\half,0,1,0,0]\   ; \  Q_4=[\half,\half,0,-1,0,0],\\Q_5&=[\half,\half,0,0,1,0]\   ;\   Q_6=[\half,\half,0,0,-1,0],\\Q_7&=[\half,\half,0,0,0,1]\   ;\   Q_8=[\half,\half,0,0,0,-1],\\Q_9&=[\half,-\half,1,0,0,0]\   ;\  Q_{10}=[\half,-\half,-1,0,0,0],\\Q_{11}&=[\half,-\half,0,1,0,0]\ ;\ Q_{12}=[\half,-\half,0,-1,0,0],\\Q_{13}&=[\half,-\half,0,0,1,0] \ ;\ Q_{14}=[\half,-\half,0,0,-1,0],\\Q_{15}&=[\half,-\half,0,0,0,1]\ ;\ Q_{16}=[\half,-\half,0,0,0,-1].\\\end{split},
\end{equation}

For the $M5$ brane, we again have $16$ supercharges. Here, we use the notation
 [$\epsilon_0,h_1,h_2,h_3,l_1,l_2$] where  $\epsilon_0$ is the$SO(2)$ charge, $h_1,h_2,h_3$ are the $SO(6)$ charges in the orthogonal basis and $l_1,l_2$ are
the $SO(5)$ charges in the orthogonal basis.
With his notation, the Qs have charges
\begin{equation}\begin{split}\label{dVIQs}
&Q_1=[\half,\half,\half,\half,\half,\half], Q_2=[\half,\half,-\half,-\half,\half,\half],\\
&Q_3=[\half,-\half,-\half,\half,\half,\half], Q_4=[\half,-\half,\half,-\half,\half,\half],\\
&Q_5=[\half,\half,\half,\half,\half,-\half],
Q_6=[\half,\half,-\half,-\half,\half,-\half],\\&Q_7=[\half,-\half,-\half,\half,\half,-\half],
Q_8=[\half,-\half,\half,-\half,\half,-\half],\\&Q_9=[\half,\half,\half,\half,-\half,\half], Q_{10}=[\half,\half,-\half,-\half,-\half,\half],\\&Q_{11}=[\half,-\half,-\half,\half,-\half,\half], Q_{12}=[\half,-\half,\half,-\half,-\half,\half],\\&Q_{13}=[\half,\half,\half,\half,-\half,-\half], Q_{14}=[\half,\half,-\half,-\half,-\half,-\half],\\&Q_{15}=[\half,-\half,-\half,\half,-\half,-\half], Q_{16}=[\half,-\half,\half,-\half,-\half,-\half].\end{split},
\end{equation}

Finally, we also specify the $d=5$ supercharges. We specify their charges in the notation[$\epsilon_0,h_1,h_2,k$], where $\epsilon_0$ is the energy, $h_1, h_2$ are the $SO(5)$ charges in the orthogonal basis and $k$ is the $Sp(2)$ R-symmetry charge.
\begin{equation}\begin{split}\label{dVQs}
&Q_1=[\half,\half,\half,\half], Q_2=[\half,\half,-\half,\half]\\&Q_3=[\half,-\half,\half,\half],
Q_4=[\half,-\half,-\half,\half]\\&Q_5=[\half,\half,\half,-\half],
Q_6=[\half,\half,-\half,-\half]\\&Q_7=[\half, -\half, \half, -\half], Q_8=[\half,-\half,-\half,-\half]
\end{split}
\end{equation}

\bibliographystyle{JHEP}
\bibliography{lgrefs}

\providecommand{\href}[2]{#2}\begingroup\raggedright\begin{thebibliography}{10}

\bibitem{Dobrev:1985qv}
V.~K. Dobrev and V.~B. Petkova, {\it All positive energy unitary irreducible
  representations of extended conformal supersymmetry},  {\em Phys. Lett.} {\bf
  B162} (1985) 127--132.

\bibitem{Dobrev:1985vh}
V.~K. Dobrev and V.~B. Petkova, {\it On the group theoretical approach to
  extended conformal supersymmetry: Classification of multiplets},  {\em Lett.
  Math. Phys.} {\bf 9} (1985) 287--298.

\bibitem{Dobrev:1985qz}
V.~K. Dobrev and V.~B. Petkova, {\it Group theoretical approach to extended
  conformal supersymmetry: Function space realizations and invariant
  differential operators},  {\em Fortschr. Phys.} {\bf 35} (1987) 537.

\bibitem{Minwalla:1997ka}
S.~Minwalla, {\it Restrictions imposed by superconformal invariance on quantum
  field theories},  {\em Adv. Theor. Math. Phys.} {\bf 2} (1998) 781--846,
  [\href{http://xxx.lanl.gov/abs/hep-th/9712074}{{\tt hep-th/9712074}}].

\bibitem{Dobrev:2002dt}
V.~K. Dobrev, {\it Positive energy unitary irreducible representations of d = 6
  conformal supersymmetry},  {\em J. Phys.} {\bf A35} (2002) 7079--7100,
  [\href{http://xxx.lanl.gov/abs/hep-th/0201076}{{\tt hep-th/0201076}}].

\bibitem{Witten:1982df}
E.~Witten, {\it Constraints on supersymmetry breaking},  {\em Nucl. Phys.} {\bf
  B202} (1982) 253.

\bibitem{Kinney:2005ej}
J.~Kinney, J.~M. Maldacena, S.~Minwalla, and S.~Raju, {\it An index for 4
  dimensional super conformal theories},  {\em Commun. Math. Phys.} {\bf 275}
  (2007) 209--254, [\href{http://xxx.lanl.gov/abs/hep-th/0510251}{{\tt
  hep-th/0510251}}].

\bibitem{Lerche:1987qk}
W.~Lerche, B.~E.~W. Nilsson, A.~N. Schellekens, and N.~P. Warner, {\it Anomaly
  cancelling terms from the elliptic genus},  {\em Nucl. Phys.} {\bf B299}
  (1988) 91.

\bibitem{Pilch:1986en}
K.~Pilch, A.~N. Schellekens, and N.~P. Warner, {\it Path integral calculation
  of string anomalies},  {\em Nucl. Phys.} {\bf B287} (1987) 362.

\bibitem{Gaiotto:2007qi}
D.~Gaiotto and X.~Yin, {\it Notes on superconformal chern-simons-matter
  theories},  {\em JHEP} {\bf 08} (2007) 056,
  [\href{http://xxx.lanl.gov/abs/0704.3740}{{\tt 0704.3740}}].

\bibitem{Cecotti:1992qh}
S.~Cecotti, P.~Fendley, K.~A. Intriligator, and C.~Vafa, {\it A new
  supersymmetric index},  {\em Nucl. Phys.} {\bf B386} (1992) 405--452,
  [\href{http://xxx.lanl.gov/abs/hep-th/9204102}{{\tt hep-th/9204102}}].

\bibitem{Denef:2007vg}
F.~Denef and G.~W. Moore, {\it Split states, entropy enigmas, holes and halos},
   \href{http://xxx.lanl.gov/abs/hep-th/0702146}{{\tt hep-th/0702146}}.

\bibitem{Sen:2007qy}
A.~Sen, {\it Black hole entropy function, attractors and precision counting of
  microstates},  \href{http://xxx.lanl.gov/abs/0708.1270}{{\tt 0708.1270}}.

\bibitem{Raju:2007uj}
S.~Raju, {\it Counting giant gravitons in ads(3)},
  \href{http://xxx.lanl.gov/abs/0709.1171}{{\tt 0709.1171}}.

\bibitem{Mack:1975je}
G.~Mack, {\it All unitary ray representations of the conformal group su(2,2)
  with positive energy},  {\em Commun. Math. Phys.} {\bf 55} (1977) 1.

\bibitem{Dolan:2002zh}
F.~A. Dolan and H.~Osborn, {\it On short and semi-short representations for
  four dimensional superconformal symmetry},  {\em Ann. Phys.} {\bf 307} (2003)
  41--89, [\href{http://xxx.lanl.gov/abs/hep-th/0209056}{{\tt
  hep-th/0209056}}].

\bibitem{Maldacena:1997re}
J.~M. Maldacena, {\it The large n limit of superconformal field theories and
  supergravity},  {\em Adv. Theor. Math. Phys.} {\bf 2} (1998) 231--252,
  [\href{http://xxx.lanl.gov/abs/hep-th/9711200}{{\tt hep-th/9711200}}].

\bibitem{Aharony:1999ti}
O.~Aharony, S.~S. Gubser, J.~M. Maldacena, H.~Ooguri, and Y.~Oz, {\it Large n
  field theories, string theory and gravity},  {\em Phys. Rept.} {\bf 323}
  (2000) 183--386, [\href{http://xxx.lanl.gov/abs/hep-th/9905111}{{\tt
  hep-th/9905111}}].

\bibitem{Bhattacharyya:2007sa}
S.~Bhattacharyya and S.~Minwalla, {\it Supersymmetric states in m5/m2 cfts},
  {\em JHEP} {\bf 12} (2007) 004,
  [\href{http://xxx.lanl.gov/abs/hep-th/0702069}{{\tt hep-th/0702069}}].

\bibitem{Gunaydin:1985tc}
M.~Gunaydin and N.~P. Warner, {\it Unitary supermultiplets of osp(8/4,r) and
  the spectrum of the s(7) compactification of eleven-dimensional
  supergravity},  {\em Nucl. Phys.} {\bf B272} (1986) 99.

\bibitem{biran1984fss}
B.~Biran, A.~Casher, F.~Englert, M.~Rooman, and P.~Spindel, {\it {The
  fluctuating seven-sphere in eleven-dimensional supergravity}},  {\em Physics
  Letters B} {\bf 134} (1984), no.~3-4 179--183.

\bibitem{Barabanschikov:2005ri}
A.~Barabanschikov, L.~Grant, L.~L. Huang, and S.~Raju, {\it The spectrum of
  yang mills on a sphere},  {\em JHEP} {\bf 01} (2006) 160,
  [\href{http://xxx.lanl.gov/abs/hep-th/0501063}{{\tt hep-th/0501063}}].

\bibitem{Seiberg:1997ax}
N.~Seiberg, {\it Notes on theories with 16 supercharges},  {\em Nucl. Phys.
  Proc. Suppl.} {\bf 67} (1998) 158--171,
  [\href{http://xxx.lanl.gov/abs/hep-th/9705117}{{\tt hep-th/9705117}}].

\bibitem{Minwalla:1998rp}
S.~Minwalla, {\it Particles on ads(4/7) and primary operators on m(2/5) brane
  worldvolumes},  {\em JHEP} {\bf 10} (1998) 002,
  [\href{http://xxx.lanl.gov/abs/hep-th/9803053}{{\tt hep-th/9803053}}].

\bibitem{Bagger:2007vi}
J.~Bagger and N.~Lambert, {\it Comments on multiple m2-branes},
  \href{http://xxx.lanl.gov/abs/0712.3738}{{\tt 0712.3738}}.

\bibitem{Sundborg:1999ue}
B.~Sundborg, {\it The hagedorn transition, deconfinement and n = 4 sym theory},
   {\em Nucl. Phys.} {\bf B573} (2000) 349--363,
  [\href{http://xxx.lanl.gov/abs/hep-th/9908001}{{\tt hep-th/9908001}}].

\bibitem{Aharony:2003sx}
O.~Aharony, J.~Marsano, S.~Minwalla, K.~Papadodimas, and M.~Van~Raamsdonk, {\it
  The hagedorn / deconfinement phase transition in weakly coupled large n gauge
  theories},  {\em Adv. Theor. Math. Phys.} {\bf 8} (2004) 603--696,
  [\href{http://xxx.lanl.gov/abs/hep-th/0310285}{{\tt hep-th/0310285}}].

\bibitem{Gross:1980he}
D.~J. Gross and E.~Witten, {\it Possible third order phase transition in the
  large n lattice gauge theory},  {\em Phys. Rev.} {\bf D21} (1980) 446--453.

\bibitem{Schnitzer:2006xz}
H.~J. Schnitzer, {\it Confinement / deconfinement transition of large n gauge
  theories in perturbation theory with n(f) fundamentals: N(f)/n finite},
  \href{http://xxx.lanl.gov/abs/hep-th/0612099}{{\tt hep-th/0612099}}.

\bibitem{Gunaydin:1984wc}
M.~Gunaydin, P.~van Nieuwenhuizen, and N.~P. Warner, {\it General construction
  of the unitary representations of anti-de sitter superalgebras and the
  spectrum of the s**4 compactification of eleven-dimensional supergravity},
  {\em Nucl. Phys.} {\bf B255} (1985) 63.

\bibitem{Claus:1997cq}
P.~Claus, R.~Kallosh, and A.~Van~Proeyen, {\it M 5-brane and superconformal
  (0,2) tensor multiplet in 6 dimensions},  {\em Nucl. Phys.} {\bf B518} (1998)
  117--150, [\href{http://xxx.lanl.gov/abs/hep-th/9711161}{{\tt
  hep-th/9711161}}].

\bibitem{Intriligator:2004xi}
K.~Intriligator and B.~Wecht, {\it Exploring the 4d superconformal zoo},
  \href{http://xxx.lanl.gov/abs/hep-th/0402084}{{\tt hep-th/0402084}}.

\bibitem{DiFrancesco:1997nk}
P.~Di~Francesco, P.~Mathieu, and D.~Senechal, {\em {Conformal field theory}}.
\newblock Springer, New York, USA, 1997.

\bibitem{fuchs1997sla}
J.~Fuchs and C.~Schweigert, {\em {Symmetries, Lie algebras and
  representations}}.
\newblock Cambridge University Press New York, NY, USA, 1997.

\bibitem{Dobrev:2004tk}
V.~K. Dobrev, {\it Characters of the positive energy uirs of d = 4 conformal
  supersymmetry},  {\em Phys. Part. Nucl.} {\bf 38} (2007) 564--609,
  [\href{http://xxx.lanl.gov/abs/hep-th/0406154}{{\tt hep-th/0406154}}].

\bibitem{kac1978rcl}
V.~Kac, {\it {Representations of classical Lie superalgebras}},  {\em Lecture
  Notes in Mathematics} {\bf 676} (1978) 597.

\bibitem{Dolan:2005wy}
F.~A. Dolan, {\it Character formulae and partition functions in higher
  dimensional conformal field theory},  {\em J. Math. Phys.} {\bf 47} (2006)
  062303, [\href{http://xxx.lanl.gov/abs/hep-th/0508031}{{\tt
  hep-th/0508031}}].

\bibitem{Bianchi:2006ti}
M.~Bianchi, F.~A. Dolan, P.~J. Heslop, and H.~Osborn, {\it N = 4 superconformal
  characters and partition functions},  {\em Nucl. Phys.} {\bf B767} (2007)
  163--226, [\href{http://xxx.lanl.gov/abs/hep-th/0609179}{{\tt
  hep-th/0609179}}].

\end{thebibliography}\endgroup
\end{document}